\documentclass[prd,showpacs
]{revtex4}

\usepackage{simplewick,graphicx}

\usepackage{amsmath,amssymb,amsfonts}
\newcommand{\RR}{{\mathbb R}}

\newcounter{mnotecount}[section]

\begin{document}

\title{Probability distributions for quantum stress tensors in 
four dimensions}

\author{Christopher J. Fewster}
\email{chris.fewster@york.ac.uk}
\affiliation{Department of Mathematics, University of York, 
Heslington, York YO10 5DD,
United Kingdom}

\author{L. H. Ford}
\email{ford@cosmos.phy.tufts.edu}
\affiliation{Institute of Cosmology, Department of Physics and
  Astronomy, 
Tufts University, Medford, Massachusetts 02155, USA}
\author{Thomas A. Roman}
\email{roman@ccsu.edu}
\affiliation{
Department of Mathematical Sciences, Central Connecticut State
University, 
New Britain, Connecticut 06050, USA}

\begin{abstract}
We treat the probability distributions for quadratic quantum fields, averaged
with a Lorentzian test function, in four-dimensional Minkowski vacuum. These 
distributions share some properties with previous results in two-dimensional
spacetime. Specifically, there is a lower bound at a finite negative value, but
no upper bound. Thus arbitrarily large positive energy density fluctuations are
possible. We are not able to give closed form expressions for the probability 
distribution, but rather use calculations of a finite number of moments to estimate
the lower bounds, the asymptotic forms for large positive argument, and possible
fits to the intermediate region. The first 65 moments are used for these purposes. 
All of our results are subject to the caveat that these distributions are not uniquely 
determined by the moments.  However, we also give bounds on the cumulative 
distribution function that are valid for any distribution fitting these moments.
We apply the asymptotic form
of the electromagnetic energy density distribution to estimate the nucleation
rates of black holes and of Boltzmann brains.
\end{abstract}

\maketitle

\section{Introduction}
\label{sec:intro}

There has been extensive work in recent decades on the definition and
use of the expectation value of a quantum stress tensor operator. When
this expectation value is used as the source in the Einstein equations,
the resulting semiclassical theory gives an approximate description of
the effects of quantum matter fields upon the gravitational field. 
This theory gives, for example, a plausible description of the
backreaction of Hawking radiation on black hole spacetimes~\cite{BD}. 

However, the semiclassical theory does not describe the effects of
quantum fluctuations of the stress tensor around its expectation
value. Quantum stress tensor fluctuations and the resulting passive
fluctuations of gravity have been the subject of several papers
in recent years~
\cite{WF01,Borgman,Stochastic,FW04,TF06,PRV09,FW07,CG97,WKF07,FMNWW10,WHFN11,LN05,Wu2007}.
However, most of these papers deal with effects described by the
correlation function of a pair of stress tensor operators, and ignore
higher-order correlation functions.  
 
One way to include these higher-order correlations is through the
probability distribution of a smeared stress tensor operator. This
distribution was given recently for Gaussian averaged
conformal stress tensors in
two-dimensional flat spacetime~\cite{FFR10}. This result will be discussed
further in Sec.~\ref{sec:SGD}. A recent attempt to define probability
distributions for quantum stress tensors in four dimensions was made
by Duplancic, {\it et al}~\cite{Duplancic}.  However, these authors attempt to
define distributions for stress tensor operators at a single spacetime
point. 
Because such operators do not have well-defined moments,
the resulting probability distribution is not well-defined. In our view,
only temporal or spacetime averages of quantum stress tensors have 
meaningful probability distributions in four dimensions. 
Furthermore, these averages should
be normal ordered, resulting in a zero mean for the vacuum probability
distribution and a nonzero probability of finding negative values.
None of these conditions are satisfied by the distribution proposed 
in Ref.~\cite{Duplancic}.

The purpose of the present paper is to obtain information about the
form of the probability distribution for averaged stress tensors in
four-dimensional spacetime from calculations of a finite set of
moments. This method was used in Ref.~\cite{FFR10} to infer the
distribution for $\varphi^2$, with Lorentzian averaging, 
where $\varphi$ is a massless scalar
field four-dimensional Minkowski spacetime. The result matches a
shifted Gamma distribution to extremely high numerical accuracy. 
 Unfortunately, the probability distribution of the smeared energy
density for massless scalar and electromagnetic fields cannot
be found so precisely. However, under certain assumptions to be
detailed later, we are able to give approximate lower bounds and asymptotic
tails for these cases, and to give a rough fit to the intermediate
part of the distribution.

An important point arises here.
Throughout this paper, all quadratic operators
are understood to be normal-ordered with respect to the Minkowski
vacuum state. However, the smeared normal ordered operators are
defined, in the first instance, only as symmetric operators on a dense
domain in the Hilbert space (assuming a real-valued smearing function) and 
it is possible that there is more than one way of extending them to provide self-adjoint operators 
\footnote{The existence of at least one self-adjoint extension is guaranteed on general grounds 
because the operators commute with complex conjugation of
the $n$-particle wavefunctions in Fock space.} 
The operators of greatest
interest to us are bounded from below on account of quantum inequalities 
(see Sect.~\ref{sec:QI}) and so there is a distinguished Friedrichs extension (see Ref.~\cite{RSii}, 
Theorem X.23),
whose lower bound coincides with the sharpest possible quantum inequality
bound. It is this operator that we have in mind when we discuss the probability
distribution of individual measurements of the smeared operator in the vacuum
state. The question of whether there is more than one self-adjoint extension, 
i.e., whether the normal ordered expressions fail to be essentially self-adjoint, 
is nontrivial and not fully resolved. Recent results (not, however, immediately applicable to our 
situation) and references may be found in Ref.~\cite{Sanders}. If there are distinct self-adjoint 
extensions, their corresponding probability distributions will all share the same moments in the 
vacuum state. 

This links to the wider issue of whether or not the moments of the probability
distribution determine the distribution uniquely. There is a rich
theory concerning this question, which is reviewed in Ref.~\cite{Simon}. As will
be discussed below, some of the moments we study grow too fast to be
covered by well-known sufficient criteria (due to Hamburger and Stieltjes) for
uniqueness. This does not prove that the distribution is nonunique (nor
would the existence of distinct self-adjoint extensions prove nonuniqueness)
and we have not been able to resolve the question of uniqueness. However,
in Sect.~\ref{sect:CDF} we prove that {\em any} probability
distribution with the moments we find has a cumulative distribution function
close to that corresponding to the fitted asymptotic tail. As various
applications (see Ref.~\cite{CMP11} and Sect.~\ref{sec:apps}) depend
only on the rough form of the tail, the possible lack of uniqueness is
not as crucial as might be thought. Further discussion of this point can be
found in Sect.~\ref{sec:unique}.

\section{Review of Some Previous Results}

Here we will briefly summarize selected aspects of two topics,
quantum inequality bounds on expectation values, and known results
for probability distributions. Both of these related topics are
important for the present paper.

\subsection{Quantum Inequalities}
\label{sec:QI}

Quantum inequalities are lower bounds on the smeared expectation
values of quantum stress tensor components in arbitrary quantum 
states~\cite{F78,F91,FR95,FR97,Flanagan97,FewsterEveson98,Fe&Ho05}.
In two-dimensional spacetime, the sampling may be over either space,
time, or both. In four dimensions, there must be a sampling either over time alone,
or over both space and time, as there are no lower bounds on purely spatially
sampled operators~\cite{FHR02}. Here we will be concerned with sampling
in time alone, in which case a quantum inequality takes the form
\begin{equation}
\int_{-\infty}^\infty f(t)\, \langle T(t,0) \rangle \, dt \geq
-\frac{C}{\tau^d} \,,
 \label{eq:QI}
\end{equation}
where $T$ is a normal-ordered quadratic operator, which is classically 
non-negative, and $f(t)$ 
is a sampling function with characteristic width $\tau$.  
Here $C$ is a numerical constant, typically small
compared to unity, and $d$ is the number of spacetime dimensions.

Although quantum field theory allows negative expectation values of the energy
density, quantum inequalities place strong constraints on the effects of this negative
energy for violating the second law of thermodynamics~\cite{F78}, maintaining
traversable wormholes~\cite{FR96} or warpdrive spacetimes~\cite{PF97}. 
The implication of Eq.~(\ref{eq:QI})
is that there is an inverse power relation between the magnitude and duration
of negative energy density. 

For a massless scalar field in two-dimensional spacetime, 
Flanagan~\cite{Flanagan97}  has found a 
formula for the constant $C$ for a given $f(t)$ which makes Eq.~(\ref{eq:QI})
an optimal inequality, and has constructed the quantum state in which the
bound is saturated. This formula is
\begin{equation}
C = \frac{1}{6 \pi} \, \int^\infty_{-\infty} du \left( \frac{d}{du} \sqrt{g(u)}\right)^2\,,
   \label{eq:Flanagan}
\end{equation}
where  $f(t)=\tau^{-1}g(u)$ and $u=t/\tau$.
This is the $c=1$ special case of a general result for unitary, positive energy, conformal field theories 
in two dimensions, 
where $c$ is the central charge, in which the left-hand side of~\eqref{eq:Flanagan} is multiplied by $c$~\cite{Fe&Ho05}.
In four-dimensional spacetime, Fewster and  Eveson~\cite{FewsterEveson98}  have
derived an analogous formula for $C$, but in this case the bound is not
necessarily optimal.

\subsection{Shifted Gamma Distributions}
\label{sec:SGD}

Here we briefly recall the main results of Ref.~\cite{FFR10}. First, we determined the
probability distribution for individual measurements, in the vacuum state, 
of the Gaussian sampled energy density 
\begin{equation}
\rho =\frac1{\sqrt{\pi}\,\tau} \int_{-\infty}^{\infty} 
T_{tt}(t,0)\, {\rm e}^{-t^2/\tau^2} \, dt
\end{equation}
of a general conformal field theory in two-dimensions. This was achieved by
finding a closed form expression for the generating function of the moments $\langle \rho^n\rangle$
of $\rho$, from which the probability distribution was obtained by inverting a Laplace transform.
The resulting distribution is conveniently expressed in terms of  the dimensionless variable
$x = \rho \, \tau^2$ and is a shifted Gamma distribution:
\begin{equation}
P(x) =\vartheta(x+x_0)
\frac{\beta^{\alpha}(x+x_0)^{\alpha-1}}{\Gamma(\alpha)} 
\exp(-\beta(x+x_0)) \,,
\label{eq:shifted_Gamma}
\end{equation}
with parameters
\begin{equation}
x_0 = \frac{c}{12\pi},\qquad \alpha = \frac{c}{12}, \qquad
\beta = \pi \,.
\end{equation}
Here $x = -x_0$ is the infimum of the support of the probability distribution, which we will 
often call the {\em lower bound of the distribution}, and $c>0$ is the central charge, which is equal to 
unity for the massless scalar field. Using the binomial theorem
and standard integrals, the $n$'th moment 
\begin{equation}
a_n = \int x^n \, P(x)\, dx \,,   \label{eq:moment} 
\end{equation}
of $P$ is easily found to be
\begin{equation}
a_n =\frac{ x_0^n}{\Gamma(\alpha)}\sum_{k=0}^n \frac{(-1)^{n-k}}{(\beta x_0)^k}
\binom{n}{k} \Gamma(k+\alpha)  = (-x_0)^n\, \,{}_2{F}{_0}(\alpha,-n; (\beta x_0)^{-1}),
\end{equation}
where ${}_2{F}{_0}$ is a generalized hypergeometric function.

The lower bound, $-x_0$, for the probability distribution for energy density
fluctuations in the vacuum for $c=1$ is exactly Flanagan's optimum lower bound,
Eq.~(\ref{eq:Flanagan}), on
the Gaussian sampled expectation value and, for all $c>0$, coincides with the result of
Ref.~\cite{Fe&Ho05}. As was argued in  Ref.~\cite{FFR10},
this is a general feature, giving a deep connection between quantum inequality
bounds and stress tensor probability distributions. The quantum inequality
bound is the lowest eigenvalue of the sampled operator, and is hence
the lowest possible expectation value and the smallest result which
can be found in a measurement. That the probability distribution for
vacuum fluctuations actually extends down to this value is more
subtle and depends upon special properties of the vacuum state. In
essence, the Reeh-Schlieder theorem implies a nonzero overlap between
the vacuum and the generalized eigenstate of the sampled operator 
with the lowest eigenvalue.

There is no upper bound on the support of $P(x)$, as arbitrarily large values of the
energy density can arise in vacuum fluctuations. Nonetheless, for the
massless scalar field, negative values are much more likely; 84\% of
the time, a measurement of the Gaussian averaged energy density will
produce a negative value. However, the positive values found the
remaining 16\% of the time will typically be much larger, and the
average [first moment of $P(x)$] will be zero.

The asymptotic positive tail of $P(x)$ has recently been used by 
Carlip {\it et al}~\cite{CMP11} to draw conclusions about the small
scale structure of spacetime in a two-dimensional model. These
authors argue that large positive energy density fluctuations tend
to focus light rays on small scales, and cause spacetime to break into
many causally disconnected domains at scales somewhat above the 
Planck length.
 
In Ref.~\cite{FFR10}, we also reported on calculations of the 
moments of ${:}\varphi^2{:}$ averaged with a Lorentzian, where $\varphi$ is a 
massless scalar field in four-dimensional spacetime. It appears that the
probability distribution is also a shifted gamma function in this case.
Define a dimensionless variable $x$ by
\begin{equation}\label{eq:xforphi2}
x = (4\pi \tau)^2 \int_{-\infty}^\infty f(t)\,  \varphi^2 \, dt\,,
\end{equation}
where
\begin{equation}
f(t) = \frac{\tau}{\pi(t^2 +\tau^2)}\,.  \label{eq:lorentzian}
\end{equation}
There is good evidence that the probability distribution is to be Eq.~(\ref{eq:shifted_Gamma}) with 
the parameters 
\begin{equation}
\alpha  =\frac{1}{72}, \qquad 
\beta    = \frac{1}{12}, \qquad
x_0 = \frac{1}{6}\, .  \label{eq:phi2-parameters}
\end{equation}
These parameters were determined empirically by fitting to the first three
calculated moments. However, the resulting distribution matches the
first sixty-five moments exactly (agreement had been checked up to the twentieth moment
at the time of writing of Ref.~\cite{FFR10}), so there can little doubt that it is correct. 
The details of this calculation are given in Sect.~\ref{sec:moments} and Appendix~\ref{appx:comp}.

Furthermore, the probability distribution for both the two-dimensional stress
tensor and the four-dimensional ${:}\varphi^2{:}$ is uniquely determined by its moments, as a
consequence of the Hamburger moment theorem~\cite{Simon}. 
This states that if 
$a_n$ is the $n$-th moment of a probability distribution $P(x)$,
then there is no other probability distribution with the same moments 
provided there exist constants $C$ and $D$ such
that
\begin{equation}
 |a_n| \leq C D^n\, n!  \label{eq:Hamburger}
\end{equation}
for all $n$. This condition is a sufficient, although not necessary, condition
for uniqueness, and is fulfilled by the moments of the shifted Gamma
distribution. The Hamburger moment theorem is also an existence
result: given a real sequence $\{a_n\}$, $n= 0,1,2,\cdots$ with $a_0=1$,
such that the $N\times N$-matrix $H^{(N)}_{mn}=a_{m+n}$
($0\le m,n\le N-1$) is strictly positive definite for every $N=1,2,\ldots$, then  there exists at least one 
associated probability distribution for which the $a_n$ are the moments.

\section{Moments and Moment Generating Functions}
\label{sec:moments}

\subsection{Explicit Calculation of Moments}
\label{sec:explicit}

In this section, we describe how the moments of a quadratic quantum
operator may be calculated explicitly. Let $\phi$ be a free quantum
field or a derivative of a free field, and let $T$ be the smeared 
normal ordered square of $\phi$:
\begin{equation}
T = \int {:}\phi^2{:}(x)\, f(x) \,dx\,,
\end{equation}
where $f$ is a sampling function. In our detailed calculations, the
smearing will be in time only, and $f = f(t)$ will be the Lorentzian
function of Eq.~(\ref{eq:lorentzian}), 
but our preliminary discussion can be more general. The
$n$-th moment $\mu_n$ of $T$ is formed by smearing the  vacuum expectation
value
\begin{equation}
G_n(x_1,\ldots,x_n) = \langle {:}\phi^2{:}(x_1)\cdots{:}\phi^2{:}(x_n)\rangle
\end{equation}
over $n$ copies of $f$. By Wick's theorem, this quantity is equal to
the sum of all contractions of the form
\begin{equation}
\contraction{}{\phi(x_1)}{\phi(x_1)}{\phi(x_2)}
\bcontraction{\phi(x_1)}{\phi(x_1)}{\phi(x_2)\phi(x_2)}{\phi(x_3)}
\contraction{\phi(x_1)\phi(x_1)\phi(x_2)}{\phi(x_2)}{\phi(x_3)\phi(x_3)\cdots\phi(x_n)}
{\phi(x_n)}
\bcontraction{\phi(x_1)\phi(x_1)\phi(x_2)\phi(x_2)\phi(x_3)}{\phi(x_3)}{\cdots}{\phi(x_n)}
\phi(x_1)\phi(x_1)\phi(x_2)\phi(x_2)\phi(x_3)\phi(x_3)\cdots\phi(x_n)\phi(x_n) \,.
\end{equation}
The contractions are subject to the rules that no $\phi(x_i)$ is contracted with the other
copy of itself and all fields are contracted, with each contraction
\begin{equation}
\contraction{}{\phi(x_i)}{\cdots}{\phi(x_j)}
\phi(x_i)\cdots\phi(x_j)\end{equation}
contributing a factor $\langle \phi(x_i)\phi(x_j)\rangle$. 

\begin{figure} 
\begin{center}
\includegraphics[height=5cm]{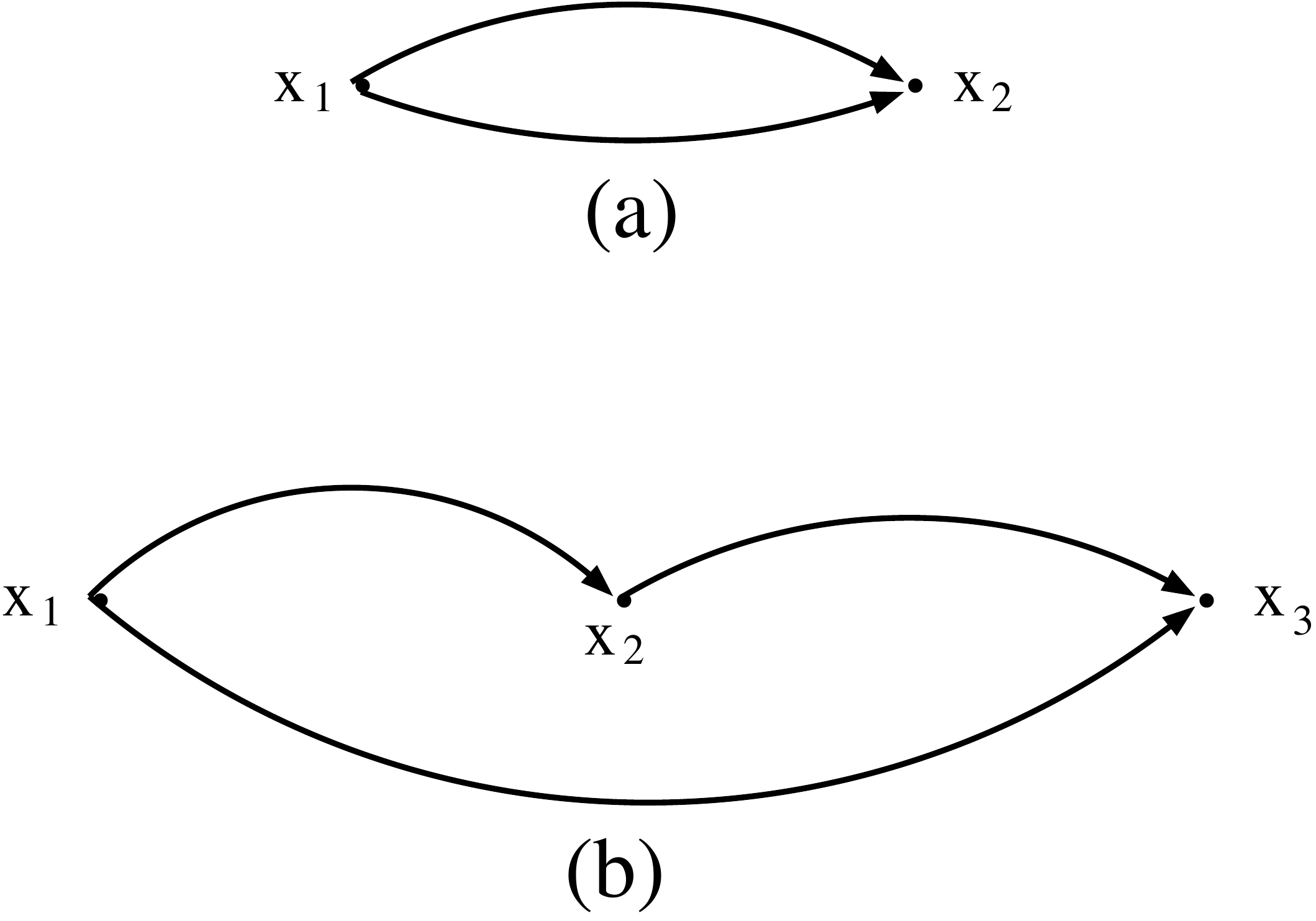}
\end{center}
\caption{The graphs for $n=2$ (a), and $n=3$ (b) are illustrated.}  
\label{fig:graph} 
\end{figure}

It is convenient to represent the contractions by graphs with $n$
vertices labelled $x_1,\ldots,x_n$ placed in order from left to right so that
(1) every vertex is met by exactly two lines; (2) every line is directed, 
pointing to the right; (3) no vertex is connected to itself by a line.
For each graph every line from $x_i$ to $x_j$ contributes the factor $\langle
\phi(x_i)\phi(x_j)\rangle$ and we supply a combinatorial
factor that gives the number of contractions represented by a given
graph; we then sum over all distinct graphs of the above type to
obtain $G_n(x_1,\ldots,x_n)$. For example, the graph in
Fig.~\ref{fig:graph}a describes the two contractions which contribute
to the second moment
\begin{equation}
\mu_2 = 2\int dx_1\,dx_2 f(x_1)f(x_2)\langle \phi(x_1)\phi(x_2)\rangle^2 \,,
\end{equation}
so the combinatorial factor for $n=2$ is $2$, while Fig.~\ref{fig:graph}b corresponds to the eight
contractions pairing a $\phi(x_1)$ with a
$\phi(x_2)$, a $\phi(x_1)$ with a $\phi(x_3)$ and a $\phi(x_2)$ with a $\phi(x_3)$, e.g.,
\begin{equation}
\contraction{\phi(x_1)}{\phi(x_1)}{}{\phi(x_2)}
\contraction{\phi(x_1)\phi(x_1)\phi(x_2)}{\phi(x_2)}{\phi(x_3)}{\phi(x_3)}
\bcontraction{}{\phi(x_1)}{\phi(x_1)\phi(x_2)\phi(x_2)}{\phi(x_3)}
\phi(x_1)\phi(x_1)\phi(x_2)\phi(x_2)\phi(x_3)\phi(x_3)
\end{equation}
and
\begin{equation}
\contraction{}{\phi(x_1)}{\phi(x_1)}{\phi(x_2)}
\contraction{\phi(x_1)\phi(x_1)\phi(x_2)}{\phi(x_2)}{}{\phi(x_3)}
\bcontraction{\phi(x_1)}{\phi(x_1)}{\phi(x_2)\phi(x_2)\phi(x_3)}{\phi(x_3)}
\phi(x_1)\phi(x_1)\phi(x_2)\phi(x_2)\phi(x_3)\phi(x_3) \,.
\end{equation}

Note that the moments $\mu_n$ have dimensions of inverse powers
of length, which depend upon the specific choice of $\phi$. It is
convenient to rescale the $\mu_n$ and define dimensionless moments
$a_n$. Our explicit calculations of moments assume the Lorentzian sampling
function of width $\tau$ given in Eq.~(\ref{eq:lorentzian}). In the case
that $\phi=\varphi$, the massless scalar field in four dimensions, we take
\begin{equation}
a_n = (4\pi \tau)^{2n} \, \mu_n \,.
\end{equation}
For the case that $\phi=\dot{\varphi}$, we take
\begin{equation}
a_n = (4\pi \tau^2)^{2n} \, \mu_n \,.        \label{eq:4d-scale}
\end{equation}
We also take the latter form for the cases of the squared electric field, and scalar and
electromagnetic field energy densities.

\subsection{Moment Generating Functions}
\label{sec:gen}

For $n\geq 4$, the Wick expansion involves both connected and
disconnected graphs. However, we need
not consider the disconnected graphs explicitly, as the moment generating function $M$
is the exponential of $W$, the generating function for the connected
graphs. The full moment generating function is defined by
\begin{equation}\label{eq:Mlambda}
M(\lambda) = \sum_{n=0}^\infty \frac{\lambda^n\;a_n}{n!} \,,
\end{equation}
so the $n$-th moment has the expression
\begin{equation}
a_n = \left(\frac{d^n M}{d \lambda^n} \right)_{\lambda = 0} \,.
\end{equation}
The connected moment generating function, $W$, has an analogous
definition, but in terms of the dimensionless connected moments $C_n$ only.
These are the moments which arise from counting only connected graphs. 
For $n=2$, there is a single connected graph, with combinatorial factor $1$ as already 
described. For $n>2$, there are $\frac{1}{2}(n-1)!$ distinct connected graphs,
each with a combinatorial factor $2^n$~\footnote{We see that each connected graph
corresponds to $2^n$ terms in the Wick expansion for $n>2$ as follows. Choose one of the lines 
emanating from $x_1$, which can correspond to any of four possible contractions. Continuing
around the diagram, each line can correspond to two possible 
contractions until the last
line, which is fixed. Thus a total of $4\times 2^{n-2}\times 1=2^n$
contractions are represented by each graph.}.  
Of course, the enumeration of
these graphs becomes rapidly unmanageable, and one must exploit further
degeneracies among the graphs to reduce the counting. For sampling
using the Lorentzian function, it is possible to reduce the number of 
terms to the number of distinct partitions of $n$ into an even number of terms. 
This grows much more slowly than $\frac{1}{2}(n-1)!$: for example,
for $n=30$, we require $2811$ terms instead of $29!/2 \approx 4.4\times 10^{30}$.
Further details can be found in Appendix~\ref{appx:comp}

Our procedure will be
to explicitly compute a finite number $N$ of connected moments, which
allows $W$ to be approximated as an $N$-th degree polynomial in
$\lambda$. 
We then use
\begin{equation}
M = {\rm e}^W   \label{eq:genfnts}
\end{equation}
to find $M$, which may also be approximated as an $N$-th degree polynomial.
Finally, the first $N$ moments $a_n$ may be read off from the
coefficients of this polynomial. We emphasize that this procedure makes sense
whether or not the series~\eqref{eq:Mlambda} converges; expressions such 
as \eqref{eq:genfnts} are simply convenient expressions for the combinatorial 
relation between different moments and may be understood as formal power series.

Consider the case of $\varphi^2$ in four dimensions, where $\varphi$ is a massless
scalar field and the average is in the time direction only. The two-point
function which appears in the integrals for the moments is now
\begin{equation}
\langle \phi(t)\phi(t')\rangle = \langle \varphi(t)\varphi(t')\rangle
= -\frac{1}{4 \pi^2(t - t' -i \epsilon)^2}=\frac{1}{4\pi^2}\int_0^\infty d\alpha\,\alpha 
{\rm e}^{-i\alpha(t-t'-i\epsilon)}  \,.
\end{equation}
The corresponding dimensionless moments were calculated 
using MAPLE for $N\le 65$, and the resulting moments
up to $N=23$ are listed in the first column of Table~\ref{table:moments}. Our computations 
were exact and give the $a_n$ as rational numbers. However, for ease
of display, the results have been rounded to five significant figures. The full set of exact moments is 
available as Supplementary Material~\cite{SM}.
As stated earlier, these moments may be used to infer that the probability
distribution of the quantity in~\eqref{eq:xforphi2} is a shifted gamma
given by Eqs.~(\ref{eq:shifted_Gamma}) and (\ref{eq:phi2-parameters}). Only the first three moments are needed for this fit, but the
result reproduces the first 65 moments exactly, a spectacular agreement.

Next we turn to the case where $\phi = \dot{\varphi}$ and calculate
several of the moments of the Lorentz-smearing of $\dot{\varphi}^2$. In this case we use
\begin{equation}
\langle \phi(t)\phi(t')\rangle =
 \langle \dot{\varphi}(t)\dot{\varphi}(t')\rangle
= \frac{3}{2 \pi^2(t - t' -i \epsilon)^4}
=\frac{1}{4\pi^2}\int_0^\infty d\alpha\,\alpha^3 e^{-i\alpha(t-t'-i\epsilon)}
 \,.   \label{eq:phi_dot_corr}
\end{equation}
As before, the moments were computed exactly as rational numbers using MAPLE for 
$N \leq 65$~\cite{SM}, and the resulting moments
up to $N=23$ are listed in the second column of Table~\ref{table:moments}. 

\begin{table}[htdp]
\caption{Lorentzian smearings of the Wick square of the free massless field $\varphi^2$,
the Wick square of its time derivative $\dot{\varphi}^2$, the square of the
electric field strength $E^2$, and the energy densities of the scalar and
electromagnetic fields $\rho_S$ and $\rho_{EM}$ respectively.}
\label{table:moments}
\begin{center}
\begin{tabular}{|c|c|c|c|c|c|} \hline
 $n$ & $\varphi^2$ &$\dot{\varphi}^2$  & $E^2$ & $\rho_S $ & $\rho_{EM}$    \\ \hline 
0 & 1  & 1 & 1 & 1 & 1  \\ \hline
1 & 0  & 0  & 0 & 0  & 0  \\ \hline
2 &  2 & 9/2 & 6 & 3/2 & 3  \\ \hline
3 & 48  & 1890 & 1680  & 525/2  & 420  \\ \hline
4 & 1740  & $2.5516 \times 10^6$  & $1.5121 \times 10^6$ & $1.6538 \times 10^5$  & $1.8903 \times 10^5$  \\ \hline
5 & 83904  & $8.5527 \times 10^9$ & $3.3789 \times 10^9$ & $2.7057 \times 10^8$  &   $2.1119 \times 10^8$ \\ \hline
6 & $5.0516 \times 10^6$ & $6.0498 \times 10^{13}$ & $1.5934 \times 10^{13}$ & $9.4918 \times 10^{11}$  & $4.9794 \times 10^{11}$ \\ \hline
7 &  $3.6472 \times 10^8$ & $7.9890 \times 10^{17}$  & $1.4027 \times 10^{17}$ & $6.2499 \times 10^{15}$ &  $2.1918 \times 10^{15}$ \\ \hline
8 & $3.0708 \times 10^{10}$  & $1.7862 \times 10^{22}$  & $2.0908 \times 10^{21}$ & $6.9804 \times 10^{19}$ & $1.6334 \times 10^{19}$  \\ \hline
9 & $2.9538 \times 10^{12}$  & $6.2613 \times 10^{26}$ & $4.8861 \times 10^{25}$ & $1.2231 \times 10^{24}$ & $1.9086 \times 10^{23}$  \\ \hline
10 & $3.1956 \times 10^{14}$  & $3.2427 \times 10^{31}$ & $1.6870 \times 10^{30}$ & $3.1669 \times 10^{28}$  &  $3.2949 \times 10^{27}$ \\ \hline
11 & $3.8406 \times 10^{16}$  & $2.3696 \times 10^{36}$ & $8.2184 \times 10^{34}$  & $1.1570 \times 10^{33}$ & $8.0257 \times 10^{31}$  \\ \hline
12 & $5.0767 \times 10^{18}$  & $2.3561 \times 10^{41}$ & $5.4477 \times 10^{39}$ & $5.7522 \times 10^{37}$ & $2.6600 \times 10^{36}$  \\ \hline
13 & $7.3196 \times 10^{20}$  & $3.0960 \times 10^{46}$ & $4.7723 \times 10^{44}$ & $3.7793 \times 10^{42}$ &  $1.1651 \times 10^{41}$ \\ \hline
14 & $1.1432 \times 10^{23}$  & $5.2487 \times 10^{51}$ & $5.3938 \times 10^{49}$ & $3.2036 \times 10^{47}$ &  $6.5843 \times 10^{45}$ \\ \hline
15 & $1.9226 \times 10^{25}$  &$1.1252 \times 10^{57}$  &$7.7085 \times 10^{54}$  & $3.4338 \times 10^{52}$ & $4.7049 \times 10^{50}$  \\ \hline
16 & $3.4641 \times 10^{27}$  & $2.9981\times 10^{62}$ & $1.3693 \times 10^{60}$ & $4.5748 \times 10^{57}$ &  $4.1789 \times 10^{55}$ \\ \hline
17 & $6.6572 \times 10^{29}$  & $9.7841 \times 10^{67}$ & $2.9791 \times 10^{65}$ & $7.4647 \times 10^{62}$ &  $4.5458 \times 10^{60}$  \\ \hline
18 & $1.3592 \times 10^{32}$  & $3.8605 \times 10^{73}$ & $7.8364 \times 10^{70}$ & $1.4726 \times 10^{68}$  &  $5.9787 \times 10^{65}$ \\ \hline
19 & $2.9384 \times 10^{34}$  & $1.8209 \times 10^{79}$ & $2.4642 \times 10^{76}$ & $3.4730 \times 10^{73}$ & $9.4000 \times 10^{70}$ \\ \hline
20 & $6.7046 \times 10^{36}$  & $1.0164 \times 10^{85}$ & $9.1702 \times 10^{81}$  & $9.6935 \times 10^{78}$ & $1.7491 \times 10^{76}$  \\ \hline
21 & $1.6103 \times 10^{39}$  & $6.6549 \times 10^{90}$ & $4.0026 \times 10^{87}$ & $3.1733 \times 10^{84}$  & $3.8172 \times 10^{81}$  \\ \hline
22 & $4.0607 \times 10^{41}$  & $5.0695 \times 10^{96}$ & $2.0327 \times 10^{93}$ & $1.2087 \times 10^{90}$ &  $9.6927 \times 10^{86}$ \\ \hline
23 & $1.0727 \times 10^{44}$  & $4.4604 \times 10^{102}$  & $1.1923 \times 10^{99}$  & $5.3172 \times 10^{95}$ &  $2.8427 \times 10^{92}$   \\ \hline

\end{tabular}
\end{center}
\end{table}

Once we have a finite set of
moments for $\dot{\varphi}^2$, we can calculate the corresponding
moments for several other operators of physical interest: we give the
examples of the energy densities for the massless scalar and electromagnetic fields,
and the squares of the electric and magnetic field strengths as particular examples. 
These all take the form 
\begin{equation}\label{eq:Agenform}
A = \int_{-\infty}^\infty dt f(t) \sum_I \alpha_I {:}\phi_I^2{:}(t,0)\,,
\end{equation}
where $\alpha_I$ are constants and the $\phi_I$ are (components of) free fields [in the sense 
that the Wick expansion is valid] with two point functions
obeying
\begin{equation}\label{eq:twoptIJ}
 c_I\delta_{IJ} \langle\dot{\varphi}(t,\mathbf{x})\dot{\varphi}(t',\mathbf{x'})\rangle_{\mathbf{x}=
 \mathbf{x'}=0}
\end{equation}
in the vacuum state, where $\varphi$ is the massless free scalar field as before
and the $c_I$ are constants.
Defining the dimensionless moments for $A$ in the 
same way as for $\dot{\varphi}^2$, one easily sees that the contribution
of any connected diagram becomes a sum over $I$ of the contributions
from each species $\phi_I$, with no cross terms mixing different species in any 
given term. Thus
\begin{equation}
C_n(A) = \sum_I (\alpha_I c_I)^n C_n(\dot{\varphi}^2)\,,
\end{equation}
from which we may infer
\begin{equation}
W(A,\lambda) =  \sum_{n=0}^\infty \frac{\lambda^n \, C_n(A)}{n!} =
\sum_I W(\dot{\varphi}^2,\alpha_I c_I \lambda)
\end{equation}
and
\begin{equation}\label{eq:Mgenform}
M(A,\lambda) = e^{W(A,\lambda)} = \prod_I M(\dot{\varphi}^2,\alpha_I c_I \lambda)\,.
\end{equation}
These results hold for arbitrary smearing on the time axis. 

This procedure may be applied to 
the energy density operator for the massless scalar field
\begin{equation}
\rho_S = \frac{1}{2} \left( \dot{\varphi}^2 +
\partial_i \varphi\, \partial^i \varphi \right)\,,
\end{equation} 
because 
\begin{align}
 \langle \dot{\varphi}(t) \partial^i \varphi(t')
\rangle_{\mathbf{x}=\mathbf{x'}=0} &= 0 \\ 
 \langle \partial_i\varphi(t)
\partial_j \varphi(t')\rangle_{\mathbf{x}=\mathbf{x'}=0}& =
\frac{1}{3}\, \delta_{ij}\, \langle \dot{\varphi}(t)
 \dot{\varphi}(t')\rangle_{\mathbf{x}=\mathbf{x'}=0} \,,
\end{align}
which is seen by direct computation of the left-hand side and comparison
with Eq.~(\ref{eq:phi_dot_corr}).   Thus we find 
\begin{equation}
C_n(\rho_S) = \left(\frac{1}{2^n} + \frac{3}{6^n}\right)
 C_n( \dot{\varphi}^2) \,;     \label{eq:conn_rel1}
\end{equation}
the factor of $3$ appearing in one of the numerators corresponds
to the spatial dimension. Thus
\begin{align}
W(\rho_S,\lambda)  & = W\left(\dot{\varphi}^2,\frac{1}{2} \lambda \right) + 
3 W\left(\dot{\varphi}^2,\frac{1}{6} \lambda \right) \\
\intertext{and}
M(\rho_S,\lambda) & = M( \dot{\varphi}^2,\frac{1}{2}\lambda) \; 
\left[ M\left(\dot{\varphi}^2,\frac{1}{6}\lambda\right) \right]^3 \,.  \label{eq:M-rho_S}
\end{align}
Again, these results should be understood as a relation between formal power series. Concretely, 
given the first $N$ moments of $ \dot{\varphi}^2$, we can approximate
$ M( \dot{\varphi}^2,\lambda)$ as a polynomial, and then use the above relation
to find the first $N$ moments of $\rho_S$. The results are tabulated in the
fourth column of Table~\ref{table:moments}.

Similarly, the components of the square of the electric and magnetic field strength $E_i$ and $B_i$ 
obey
\begin{equation}
 \langle E_i(t) \,E_j(t')\rangle_{\mathbf{x}=\mathbf{x'}=0} =
 \langle B_i(t) \,B_j(t')\rangle_{\mathbf{x}=\mathbf{x'}=0} =
\frac{2}{3}\, \delta_{ij}\, \langle \dot{\varphi}(t)
 \dot{\varphi}(t')\rangle_{\mathbf{x}=\mathbf{x'}=0} \,.
\end{equation}
Following the same line of reasoning as before, we find
\begin{equation}
 C_n(E^2) = C_n(B^2) = 3 \left(\frac{2}{3}\right)^n\, C_n( \dot{\varphi}^2)\,, 
\end{equation}
\begin{equation}
 W(E^2,\lambda) = W(B^2,\lambda) = 3 W\left(\dot{\varphi}^2,\frac{2}{3} \lambda\right)\,,
\end{equation}
and
\begin{equation}
M(E^2,\lambda) = M(B^2,\lambda) = \left[ M\left(\dot{\varphi}^2,\frac{2}{3}\lambda \right) \right]^3 \,. 
   \label{eq:M-E2}
\end{equation}
This result leads to the moments of the squared electric field, tabulated
in the third column in Table~\ref{table:moments}. The results for
the square of the magnetic field are identical. 

Finally, because we also have
\begin{equation}
 \langle E_i(t,\mathbf{x})  B_j(t',\mathbf{x'})
\rangle_{\mathbf{x}=\mathbf{x'}=0} = 0\,,
\end{equation}
the energy density of the electromagnetic field 
\begin{equation}
\rho_{EM} = \frac{1}{2} \left( E^2 + B^2 \right) 
\end{equation}
has connected moments
\begin{equation}
C_n(\rho_{EM}) = 2 \left(\frac{1}{2}\right)^n\, C_n( E^2) = 
6 \left(\frac{1}{3}\right)^n\, C_n( \dot{\varphi}^2)\,,      \label{eq:conn_rel2}
\end{equation}
and hence
\begin{equation}
W(\rho_{EM},\lambda ) =       2  W\left(E^2,\frac{1}{2}\lambda \right)
= 6 W\left(\dot{\varphi}^2,\frac{1}{3}\lambda \right) \,,
\end{equation}
and
\begin{equation}
M(\rho_{EM},\lambda ) =       \left[M\left(E^2,\frac{1}{2}\lambda \right)\right]^2 =
\left[M\left(\dot{\varphi}^2,\frac{1}{3}\lambda \right)\right]^6 
\,,
\label{eq:M-rho_EM}
\end{equation}
leading to the remaining entries in Table~\ref{table:moments}.

 An important observation is that these
moments (apart from those of the Wick square) grow too rapidly to satisfy the Hamburger moment 
criterion, 
Eq.~(\ref{eq:Hamburger}). This may be confirmed by noting that in all
cases $\ln \, a_n$ grows faster with increasing $n$ than $ n\, \ln n + c_1 n + c_0$
for any constants $c_0$ and $c_1$. In fact, the growth for $\dot{\varphi}^2$ is shown in 
Appendix~\ref{appx:asymp}
to be of the form
\begin{equation}
a_n \sim C \, D^n \, (3n-4)! \,,  \label{eq:asymp}
\end{equation}
where the constant $D$ is proved to lie in the range $3.221667 < D  < 3.616898$ (our numerical evidence suggests $D\sim 3.3586$). For probability distributions
known to be confined to a half-line, which is the case here, there is a sufficient
condition for uniqueness which is weaker than the  Hamburger moment criterion.
This is the Stieltjes criterion~\cite{Simon}, which is
\begin{equation}
a_n \leq C \, D^n \, (2n)! \,.    \label{eq:Stieltjes}
\end{equation}
Unfortunately, this criterion is also not fulfilled here.
This means that we cannot be guaranteed
of finding  a unique probability distribution $P(x)$ from these moments. This issue will be
discussed further in Sec.~\ref{sec:unique}. 

Note that in four dimensions, the
operators $(\dot{\varphi}^2$, $E^2$, $\rho_S$, and $\rho_{EM})$ all have
dimensions of $length^{-4}$. Their probability distributions $P(x)$ will be 
taken to be functions of the dimensionless variable  [See Eq.~(\ref{eq:4d-scale}).]
\begin{equation}
x = (4\pi \, \tau^2)^2 \, A\,,
\end{equation}
where $A$ is the Lorentzian time average of $(\dot{\varphi}^2$, $E^2$, $\rho_S$, 
 $\rho_{EM})$.

\subsection{Lower Bounds}
\label{sec:lower}

In general, we may use relations between different moment generating functions
to find relations between the corresponding probability distributions, and
especially between the lower bounds of these distributions. (Strictly, these are
the infima of the support of the distributions.) Let $p(x)$ and
$q(x)$ be two probability distributions, with moment generating
functions $M(p,\lambda)$ and $M(q,\lambda)$, respectively. These generating
functions can be expressed in terms of the bilateral Laplace transforms of their probability distributions:
 \begin{equation}
M(p,\lambda) = \int_{-\infty}^\infty p(x)\, {\rm e}^{\lambda x} \, dx
\end{equation}
and 
\begin{equation}
M(q,\lambda) = \int_{-\infty}^\infty q(x)\, {\rm e}^{\lambda x} \, dx \,.
\end{equation}
These integrals are guaranteed to converge at the lower limits, due to
the lower bounds on the support of our probability distributions. To assure
convergence at the upper limit, we may assume ${\rm Re }\, \lambda < 0$.
However, many of our arguments below do not require convergence
of the integrals, which may be regarded as formal power series in $\lambda$
on replacing the exponential by its Taylor series.
Now let $p*q(x)$ be a probability distribution defined as the convolution
of $p$ and $q$:
\begin{equation}
p*q(x) = \int_{-\infty}^\infty dx' \, p(x-x')\, q(x') \,. \label{eq:convol}
\end{equation}
As is well-known in probability theory, this is the distribution for the random variable obtained as
the sum of independent random variables with distributions $p$ and $q$, and 
its moment generating function is
\begin{eqnarray}
M(p*q,\lambda) &=& 
\int_{-\infty}^\infty dx\, \int_{-\infty}^\infty dx'\, p(x-x')\, q(x') \,  {\rm e}^{\lambda x} 
\nonumber \\
&=&  \int_{-\infty}^\infty dx'\, \left[\int_{-\infty}^\infty dx\,p(x-x')\,  {\rm e}^{\lambda (x-x')}
\right] q(x') \,  {\rm e}^{\lambda x'}  \nonumber \\
&=& \left[\int_{-\infty}^\infty du \, p(u)\,{\rm e}^{\lambda u}\right]\;
  \left[\int_{-\infty}^\infty dx' \, q(x')\,{\rm e}^{\lambda x'}\right]\
 = M(p,\lambda)\; M(q,\lambda)  \,,
\end{eqnarray}
where $u = x - x'$\,. Thus the moment generating function of a convolution 
is the product of the individual generating functions; again, this holds in the 
sense of formal power series, irrespective of convergence issues.

We can also give the relation of the lower bounds. As is also well-known in
probability theory, the support of a convolution $p*q$ of two distributions consists of
all values expressible as the sum of a value in the support of $p$ and a value in the support of $q$. 
In particular, the greatest lower bound on the support is the sum of the lower
bounds of the individual distributions. Explicitly, if  $b_p$ and $b_q$ be
the lower bounds of $p$ and $q$, then
\begin{equation}
p(x) = 0 \quad {\rm if} \quad x < b_p\;;\qquad q(x) = 0 \quad {\rm if} \quad x < b_q \,.
\end{equation}
The integrand of Eq.~(\ref{eq:convol}) vanishes if either $x' < b_q$, or $x-x' < b_p$.
This implies that 
\begin{equation}
p*q(x) = 0 \quad {\rm if} \quad x < b_p + b_q \,,
\end{equation}
and in fact this is the greatest lower bound. Thus the lower bound of $p*q$ is the sum of the 
bounds of $p$ and of $q$.

Next consider the effect of a rescaling of $\lambda$, and let $p_\alpha(x)=
|\alpha|\, p(\alpha\,x)$, where $\alpha \neq 0$. Then
 \begin{equation}
M(p_\alpha,\lambda) = 
|\alpha|\int_{-\infty}^\infty p(\alpha x)\, {\rm e}^{\lambda x} \, dx
= \int_{-\infty}^\infty p(x')\, {\rm e}^{(\lambda/\alpha) x'} \, dx'
= M\left(p, \frac{\lambda}{\alpha}\right) \,.
\end{equation}
Provided $\alpha>0$, $p_\alpha = 0$ if $x < b_p/\alpha$, so the effect of rescaling
$\lambda$ in $M$ is a rescaling of the lower bound by the same factor. If
$\alpha<0$, the lower bound on the support of $p_\alpha$ is $-|\alpha|^{-1}$
times the upper bound on the support of $p$, if this exists; if there is no
upper bound  on the support of $p$, then evidently $p_\alpha$ has
no lower bound in this case.

Now we may combine these results to relate the lower bounds of various 
probability distributions to that for $\dot{\varphi}^2$. Applied to
a general operator of the form~\eqref{eq:Agenform}, they suggest that
the probability distribution for $A$ is a convolution of several copies of 
the probability distribution for $\dot{\varphi}^2$, with various scalings.
For example, Eq.~(\ref{eq:M-rho_S}) suggests that the probability distribution for 
the energy density $\rho_S$, smeared along the time axis, is equal
to the convolution of four copies of the probability distribution for $\dot{\varphi}^2$, with
various scalings. In particular, recalling that $x_0(A)$ denotes the greatest lower
bound on the support of the distribution for $A$ smeared in time, this suggests that
\begin{equation}
x_0(A) = \left( \sum_I \alpha_I c_I\right) x_0(\dot{\varphi}^2)
\end{equation}
Hence Eq.~(\ref{eq:M-rho_S}) suggests that
$x_0(\rho_S) = (1/2+ 3 \times \,1/6) x_0(\dot{\varphi}^2) =
 x_0(\dot{\varphi}^2)$.  Similarly,  Eq.~(\ref{eq:M-E2}) suggests that
 $x_0(E^2) =  3 \times \,(2/3) x_0(\dot{\varphi}^2) =2\, x_0(\dot{\varphi}^2)$,
 and Eq.~(\ref{eq:M-rho_EM}) suggests that $x_0(\rho_{EM}) = 2 \times \,(1/2) x_0(E^2)
 =x_0(E^2)$. In summary,
\begin{equation}
x_0(\rho_{EM}) = x_0(E^2) = 2\, x_0(\rho_S) = 2\,  x_0(\dot{\varphi}^2) \,.
 \label{eq:bounds}
\end{equation}
Likewise, if we consider a combination such as the pressure $T_{11} = 
\frac{1}{2}(\dot{\varphi}^2 +(\partial_1\varphi)^2 - (\partial_2\varphi)^2
-(\partial_3\varphi)^2)$, we obtain the expected result that the probability
distribution is unbounded both from above and below. The above derivations should be take as 
suggestive, rather than rigorous proofs, because of concerns
about the uniqueness of the underlying probability distributions. However, it would be possible
to prove them by writing the smeared operator for $\rho_S$, for example, as a sum of mutually
commuting self-adjoint operators, each of which was essentially a multiple of the smeared 
$\dot{\varphi}^2$ operator (under a suitable unitary transformation). 
This could be done by writing the field in a basis of spherical harmonics,
in this framework, the three powers of $M(\dot{\varphi}^2,\lambda/6)$ arise from the  $\ell=1$ 
angular momentum sector, while the single power of $M(\dot{\varphi}^2,\lambda/2)$ arises from the 
$\ell=0$ sector. Indeed, one of the first quantum inequality bounds on the expectation value of 
$\rho_S$ used precisely this decomposition~\cite{FR95}. More generally, Eq.~\eqref{eq:twoptIJ} 
could be used in conjunction with Wick's theorem to show that timelike smearings of ${:}\phi_I^2{:}$ 
and ${:}\phi_J^2{:}$ commute for $I\neq J$, at least
in matrix elements between states obtained from the vacuum by applying
polynomials of smeared fields, and might be used to put the other relationships above on a firmer 
footing; we will not pursue this here.

\section{Lower Bound Estimates}
\label{sec:estimates}

Here we will discuss a technique, a Stieltjes moment test,
by which knowledge of a finite
number of moments may be used to obtain an approximate estimate of
the lower bound. If $P(x)$ is a probability distribution with a
lower bound at $x=-x_0$, then its moments are
\begin{equation}
a_n = \int_{-x_0}^\infty x^n \, P(x)\, dx \,.   \label{eq:moment2} 
\end{equation}
Let 
\begin{equation}
I(y) =  \int_{-x_0}^\infty (x +y) \,|q(x)|^2 P(x)\, dx \,,
\label{eq:I}
\end{equation}
where $q(x)$ is a polynomial and $y \geq x_0$. We see that 
$I(y) \geq0$ because the integrand in Eq.~(\ref{eq:I}) is
non-negative. If
\begin{equation}
q(x) = \sum_{n=0}^{N-1} \beta_n\, x^n \,,
\end{equation}
then
\begin{equation}
I(y) = \sum_{m,n=0}^{N-1} M_{mn}(N,y) \beta_m^* \beta_n \geq 0 \, ,
\label{eq:pos}
\end{equation}
where $M(N,y)$ is a real symmetric $N \times N$ matrix with elements
\begin{equation}
 M_{mn}(N,y) = a_{m+n+1} +y a_{m+n} \qquad (0\le m,n\le N-1)\,.
\end{equation}

 Let $\beta_n$ be the components of an eigenvector with
eigenvalue $\lambda$, then $\sum_{n=0}^{N-1}  M_{mn}(N,y) \, \beta_n
= \lambda\, \beta_m$, and $I(y) = \lambda \, \sum_{m=0}^{N-1} |\beta_m|^2$. 
It follows that $M(N,y)$ has no negative eigenvalues, that is, it
is a positive semidefinite matrix, which we denote by $M(N,y) \geq 0$.
This holds for all $N$ and all $y \geq x_0$. However, as $y$ decreases
below $x_0$, the lowest eigenvalue is eventually zero and then 
negative eigenvalues can occur.
Define $y_N$ as the minimum value of $y$ at which  $M(N,y) \geq 0$;
in practice, it is easiest to compute $y_N$ as the largest root of the 
$N$'th degree polynomial equation   
\begin{equation}
{\rm det} M(N,y) =  0 \,.
\end{equation}
Because  $M(N,y)$ is a leading principal minor of  $M(N+1,y)$, 
$M(N+1,y) \geq 0$ implies that $M(N,y) \geq 0$. Consequently,
$y_{N+1} \geq y_N$ and the sequence in $N$ converges to a limit
with
\begin{equation}
y_\infty = \lim_{N\rightarrow \infty} y_N \leq x_0 \,.
\end{equation}

Given a set of moments $a_n$ of an unknown probability distribution, we may form the 
matrices $M(N,y)$ as above and determine the values of $y_N$. The above argument shows 
that if $y_N\to\infty$ then
the $a_n$ {\em cannot} be the moments of a probability distribution whose support is bounded 
from below. On the other hand, suppose that a finite limit $y_\infty$ exists. Then for any 
probability distribution 
$\tilde{P}$ with the same moments and support bounded below by $-\tilde{x}_0$, we have 
$y_\infty\le \tilde{x}_0$. In particular, there is no probability distribution accounting for the given 
moments with support contained in $(-y_\infty,\infty)$. 

\begin{table}[t]
\caption{Table of the lower bounds, $y_N$, for both $\varphi^2$ and
$\dot{\varphi}^2$.  }
\label{tab:bounds}
\begin{center}
\begin{tabular}{|c|c|c|c|c|c|} \hline
 $N$ & $y_N ( \varphi^2)$ &$y_N (\dot{\varphi}^2)$    \\ \hline 
2  &  0.08304597359  &  0.01071401240  \\ \hline 
3  &  0.11085528820  &  0.01414254029  \\ \hline 
4  &  0.12478398360  &  0.01584995314  \\ \hline 
5  &  0.13314891433  &  0.01690199565  \\ \hline 
6  &  0.13872875370  &  0.01762865715  \\ \hline 
7  &  0.14271593142  &  0.01816742316  \\ \hline 
8  &  0.14570717836  &  0.01858660399  \\ \hline 
9  &  0.14803421582  &  0.01892432539  \\ \hline 
10  &  0.14989616852  &  0.01920370321  \\ \hline 
11  &  0.15141979779  &  0.01943965011  \\ \hline 
12  &  0.15268963564  &  0.01964226267  \\ \hline 
\end{tabular}\quad
\begin{tabular}{|c|c|c|c|c|c|} \hline
 $N$ & $y_N ( \varphi^2)$ &$y_N (\dot{\varphi}^2)$    \\ \hline
13  &  0.15376421805  &  0.01981864633  \\ \hline 
14  &  0.15468536476  &  0.01997396248  \\ \hline 
15  &  0.15548374872  &  0.02011206075  \\ \hline 
16  &  0.15618237796  &  0.02023587746  \\ \hline 
17  &  0.15679884907  &  0.02034769569  \\ \hline 
18  &  0.15734684979  &  0.02044932047  \\ \hline 
19  &  0.15783718730  &  0.02054219985  \\ \hline 
20  &  0.15827850807  &  0.02062751059  \\ \hline 
21  &  0.15867781217  &  0.02070622001  \\ \hline 
22  &  0.15904082736  &  0.02077913144  \\ \hline 
23  &  0.15937228553  &  0.02084691828  \\ \hline 
\end{tabular}\quad
\begin{tabular}{|c|c|c|c|c|c|} \hline
 $N$ & $y_N ( \varphi^2)$ &$y_N (\dot{\varphi}^2)$    \\ \hline

24  &  0.15967613018  &  0.02091014970  \\ \hline 
25  &  0.15995567400  &  0.02096931050  \\ \hline 
26  &  0.16021372020  &  0.02102481644  \\ \hline 
27  &  0.16045265677  &  0.02107702642  \\ \hline 
28  &  0.16067453067  &  0.02112625203  \\ \hline 
29  &  0.16088110659  &  0.02117276528  \\ \hline 
30  &  0.16107391397  &  0.02121680481  \\ \hline 
31  &  0.16125428495  &  0.02125858099  \\ \hline 
32  &  0.16142338519  &  0.02129828002  \\ \hline 
&& \\  && \\ \hline 
\end{tabular}
\end{center}
\end{table}

Let us first apply this method to the case of the $\varphi^2$ distribution,
given by Eqs.~(\ref{eq:shifted_Gamma}) and (\ref{eq:phi2-parameters}), 
for which the exact lower bound is known. The results of the calculation
of  the $y_N$ through $N=32$ are given in Table~\ref{tab:bounds} (computations
were performed in MAPLE to 40 digit accuracy; the reported rounded figures are stable 
under increase
of the number of digits). We can improve the estimate of the lower bound by extrapolation. A
trial function of the form $y_N = a + b/N + c/N^2$ and a least-squares fit using 
MAPLE~\footnote{Fitting
was performed using the {\tt Statistics[Fit]} command, working to 40 digit accuracy.} to
determine values of $a$, $b$ and $c$ leads to 
\begin{equation}
y_N(\varphi^2) \approx 0.166666666057-\frac{0.167821368174}{N}+\frac{0.001164170336}{N^2}
\end{equation}
The above fit was obtained using the data points for $21\le N\le 33$, with residuals of order
$10^{-12}$ over these values, and no more than $1.1\times 10^{-6}$ for $2\le N\le 20$. Using the fit 
displayed above, our lower bound estimate now becomes $y_\infty = 0.166666666057$, in extremely 
good agreement with the exact bound, $x_0 = 1/6$, obtained from
Eq.~\eqref{eq:phi2-parameters}. This suggests the conjecture that $-y_\infty$ might
also coincide with the lower bound of the probability distribution in other cases as well, but note the
caveat at the end of this section. 

A different numerical approach is to use an accelerated convergence trick: given any
sequence $y = (y_N)$, define a new sequence $L^{(k)}y$ with terms
\begin{equation}
(L^{(k)}y)_N = \frac{N+1}{k}(y_{N+1}-y_N) + y_N;
\end{equation}
for finite sequences, $L^{(k)}y$ is one term shorter than $y$. 
This is a linear map on sequences, preserving constants and acting on $y_N=1/N^p$ by
\begin{equation}
(L^{(k)}y)_N = \frac{1-p/k}{N^p} + O(1/N^{p+1})
\end{equation}
for any $p,k>0$. Thus if $y_N = a+ b N^{-k} + c N^{-\ell} + \cdots$, with $\ell>k$, 
the sequence $L^{(k)}y$ converges to $a$ as $O(N^{-\min\{\ell, k+1\}})$, rather than 
$O(N^{-k})$. 
This trick may be repeated: in the situation above, $L^{(2)}L^{(1)}y(\varphi^2)_N$ would be 
expected to converge with $O(N^{-3})$ speed to the limit. The results give values differing from 
$1/6$ by less than $10^{-6}$ for all $11\le N\le 31$. Part of the  `accelerated'  sequence is 
given in Table~\ref{tab:accelerated}. 

\begin{table}[t]
\caption{Table of the accelerated lower bounds for both $\varphi^2$ and
$\dot{\varphi}^2$.  }
\label{tab:accelerated}
\begin{center}
\begin{tabular}{|c|c|c|c|c|c|} \hline
 $N$ & $L^{(2)}L^{(1)} y_N ( \varphi^2)$ & 
$L^{(3/2)}L^{(1)}L^{(1/2)}y_N (\dot{\varphi}^2)$    \\ \hline 
21  &  0.16666653954  &  0.02361472123  \\ \hline 
22  &  0.16666655611  &  0.02361451051  \\ \hline 
23  &  0.16666656993  &  0.02361432088  \\ \hline 
24  &  0.16666658153  &  0.02361414978  \\ \hline 
25  &  0.16666659135  &  0.02361399500  \\ \hline 
26  &  0.16666659972  &  0.02361385460  \\ \hline 
27  &  0.16666660689  &  0.02361372693  \\ \hline 
28  &  0.16666661307  &  0.02361361053  \\ \hline 
29  &  0.16666661843  &  0.02361350416  \\ \hline 
30  &  0.16666662310  &  0.02361340672  \\ \hline 
31  &  0.16666662718  &    \\ \hline 
\end{tabular}
\end{center}
\end{table}

We may now apply the same procedure to the case of  $\dot{\varphi}^2$,
where the exact bound is not known. The  $y_N(\dot{\varphi}^2)$ are
also given in Table~\ref{tab:bounds}, and clearly converge more slowly than those of the  
$y_N(\varphi^2)$. Indeed, successive differences $y_{N+1}(\dot{\varphi}^2)- y_{N}(\dot{\varphi}^2)$ 
appear to decay as $O(N^{-3/2})$. A least squares fit to the trial function 
$y_N(\dot{\varphi}^2) = a + b/N^{1/2} + c/N + d/N^{3/2}$ gives
\begin{equation}
y_N(\dot{\varphi}^2)\approx
0.0236174942666 -\frac{0.012425890959}{N^{1/2}} 
-\frac{0.002768353926}{N}-
\frac{0.006533917931}{N^{3/2}}
\end{equation}
using $21\le N\le 33$, with residuals less than $1.2\times 10^{-10}$ on these values, 
and no more than
$10^{-5}$ on $6\le N\le 20$. Applying the acceleration technique, 
$L^{(3/2)}L^{(1)}L^{(1/2)} y(\varphi^2)_N$ gives a sequence differing from $0.02361$ by no more 
than $8.1\times 10^{-6}$
on $11\le N\le 30$. Taking this together with the least squares fit gives reasonable confidence in an 
estimate $y_\infty(\dot{\varphi}^2) =0.02361\pm 1\times 10^{-5}$.

%
%
%
%
%
%

In contrast, the non-optimal bound for $\dot{\varphi}^2$ and $\rho_S$, given by 
the method of Fewster and
Eveson~\cite{FewsterEveson98}, is $x_0(FE) = 27/128 \approx 0.21$, which is an
order of magnitude larger. [This bound is given by minus the right hand side of
Eq.~(5.6) in Ref.~\cite{FewsterEveson98} multiplied by $(4 \pi \tau^2)^2$.]
 If, in fact, $-y_\infty$ coincides with the lower bound of the
probability distribution, we can now use the results in
Eq.~(\ref{eq:bounds}) to write our estimates of the probability
distribution lower bounds as
\begin{equation}
-x_0(\rho_{EM}) = -x_0(E^2) \approx  -0.0472 \qquad  
-x_0(\rho_S) =- x_0(\dot{\varphi}^2) \approx -0.0236\,.
 \label{eq:bounds2}
\end{equation}
These are also estimates of the optimal quantum inequality bounds
for each field.

There is an important caveat to this reasoning, however. If the moments do not correspond
to a unique probability distribution (i.e., if it they are indeterminate in the Hamburger sense) 
then there will exist probability distributions, called {\em von Neumann solutions} in 
Ref.~\cite{Simon}, with the given moments that are pure point measures, in contrast to the 
continuum probability distribution that would be expected for the quantum field theory 
operators we study (and which we find for the $\phi^2$ case). As the moments arise from a
probability distribution supported in a half-line, there is a distinguished von Neumann solution, 
called the {\em Friedrichs solution} in Ref.~\cite{Simon}, that is supported in a 
half-line $[-x_F,\infty)$
and has the property that no other solution to the moment problem can also be supported in
$[-x_F,\infty)$. (See Appendix~C1 of~\cite{Simon} for a brief summary.) Hence if operator $A$ 
has Hamburger-indeterminate moments, we would have
$y_\infty(A)= x_F(A) < x_0(A)$. Nonetheless, the results in Eq.~\eqref{eq:bounds2} would still be true 
if the approximation signs are replaced by $\alt$. 

It is of interest to note that the magnitudes of the dimensionless lower bounds, given in
Eq.~(\ref{eq:bounds2}) are small compared to unity. Given that the probability distribution must
have a unit zeroth moment and a vanishing first moment, this implies that $P(x) \gg 1$ in
at least part of the interval $-x_0 < x <0$. Thus the spike at the lower bound found in the
two-dimensional case may be a generic feature. The small magnitudes of $x_0(\rho_{S})$
and  $x_0(\rho_{EM})$ imply strong constraints on the magnitude of negative energy which
can arise either as an expectation value in an arbitrary state, or as a fluctuation in the  
vacuum. They also imply that an individual measurement of the sampled energy density in the 
vacuum state is very likely to yield a negative value.

\section{Fits for the Approximate Form of the Probability Distribution}
\label{sec:fits}

In this section, we explore the extent to which knowledge of a finite
set of moments may be used to obtain information about $P(x)$
beyond the lower bounds found in Sec.~\ref{sec:estimates}. 

\subsection{A procedure to find the parameters of the tail of $P(x)$}
\label{sec:procedure}
 
We begin with the large $x$ limit. Let us adopt the ansatz that: 
\begin{equation}
P(x) \sim c_0 \,x^b \, e^{-a x^c} \,,
\label{eq:pofx}
\end{equation}
for large $x$. We assume that we can use this form of the tail to compute the large $n$ moments, 
and find
\begin{eqnarray}
a_n &=& \int_{-x_0}^\infty \, x^n \,P(x) \,dx \approx c_0 \, \int_0^\infty \,x^{n+b} \, e^{-a x^c}   
\nonumber \\
&=& \frac{c_0}{c} \, a^{-(n+b+1)/c} \, \, [ (n+b+1)/c -1]!  \,,
\label{eq:mun}
\end{eqnarray}
for $n \gg 1$. We expect the dominant contribution to come from $x \gg
1$, so we set the lower limit in the second integral to zero for convenience.
 
 Next we compare Eq. ~(\ref{eq:mun}) with the Eq.~(\ref{eq:asymp}) for the large
 $n$ form of the moments. This comparison reveals that we should have
 \begin{equation}
 c = \frac{1}{3}\,, \quad b =-2\, \quad a=D^{-1/3}\,,  \quad c_0 = CD/3 \,.
 \label{eq:ac0}
 \end{equation}
 With these values for $c$ and $b$, the ratio of successive moments from Eq.~(\ref{eq:mun})
 becomes 
\begin{equation}
\frac{a_{n+1}} {a_n} \approx 
\frac{3 ( n -1) (3  n -2) (3 n -1)}{a^3} \,.
\end{equation}
Now we may use the computed values of two successive moments, such as $n=64$
and $n=65$, to find the value of $a$, and then the value of $c_0$ from Eq.~(\ref{eq:mun}).
The results for the different operators are listed in Table~\ref{cvalues}. It should be 
noted that knowledge for further moments beyond $n=65$ could change the values in
this table. A rough error analysis suggests that these values are correct to about five
significant figures.

\begin{table}[htdp]
\caption{Values of the Parameters for the Tails, in the form of Eq.~(\ref{eq:pofx}).}
\label{tab:tail}
\begin{center}
\begin{tabular}{|c|c|c|c|c|} \hline
Operator & $c_0$ & $a$ & $b$ &$c$ \\ \hline 
$\dot{\varphi}^2$ & 0.47769605  & 0.6677494904 &  -2 & 1/3 \\ \hline 
 $E^2$ & 0.95539211 & 0.7643823521 & -2 & 1/3 \\ \hline
 $\rho_S$  & 0.23884802 & 0.8413116390 &  -2 &1/3 \\ \hline
 $\rho_{EM}$ & 0.95539211  & 0.9630614156 &  -2 & 1/3 \\ \hline
\end{tabular}
\end{center}
\label{cvalues}
\end{table}

The values of the constants $a$ and $c_0$ for the various cases can 
be related to one another by means of the relations between the connected 
moments, Eqs.~(\ref{eq:conn_rel1}) and (\ref{eq:conn_rel2}), derived in
Sect.~\ref{sec:gen}. First, we need the fact that the connected moments and the 
full moments rapidly approach one another for large $n$, specifically
\begin{equation}
C_n \sim a_n (1 + O(n^{-4}))\,, \quad n \gg1 \,.
\end{equation}
This relation may be demonstrated analytically, or inferred numerically from
the computed moments. This means that Eqs.~(\ref{eq:conn_rel1}) and 
(\ref{eq:conn_rel2}) hold for the full moments, $a_n$ when $n$ is large.
The former relation may be simplified to $a_n(\rho_S) \sim 2^{-n} a_n( \dot{\varphi}^2)$.
The asymptotic form, Eq.~(\ref{eq:asymp}), for the moments of $\dot{\varphi}^2$ also
holds for the other operators, but with different choices of the constants $C$ and
$D$:
\begin{equation}
C(\dot{\varphi}^2) = C(\rho_S) = \frac{1}{3} C(E^2) = \frac{1}{6} C(\rho_{EM}) \quad
{\rm and} \quad D(\dot{\varphi}^2) = 2 D(\rho_S) = \frac{3}{2} D(E^2) = 3 D(\rho_{EM}) \,. 
\end{equation}
For example,  $a_n(\rho_{EM}) =  6\, (1/3)^n\, a_n( \dot{\varphi}^2)$, 
from Eq.~(\ref{eq:conn_rel2}), implies
the above relations between $C(\dot{\varphi}^2)$ and $C(\rho_{EM})$ and between
 $D(\dot{\varphi}^2)$ and $D(\rho_{EM})$.
These relations and Eq.~(\ref{eq:ac0}) imply that
\begin{equation}
c_0(\dot{\varphi}^2) = 2 c_0(\rho_S) = \frac{1}{2} c_0(E^2) = \frac{1}{2} c_0(\rho_{EM}) \quad
{\rm and}  \quad   a(\dot{\varphi}^2) = 2^{-\frac{1}{3}}\, a(\rho_S) = 
 \left(\frac{2}{3}\right)^\frac{1}{3}\, a(E^2) = 3^{-\frac{1}{3}}\,  a(\rho_{EM}) \,. 
\end{equation}
These relations are borne out by the values in Table~\ref{cvalues}.

\subsection{Estimating when our tail fit is a good approximation}
\label{sec:estimate}

Since the Hamburger and Stieltjes moment conditions are 
not fulfilled for our moments, 
we do not know whether our probability distributions are
unique. However, if we assume that they are, then we can estimate
 the range in $x$ where we expect our fitted tails to give a good 
estimate of the actual distributions. Our general form for the 
tails of the probability distributions is approximately
\begin{equation}
P_{\rm fit}(x) \sim c_0 \,x^{-2} \, e^{-a x^{1/3}} \,.
\label{eq: pb2}
\end{equation}
As an example, for $\rho_{EM}$, 
this gives a good fit $(\leq 10 \%)$ for $n=4,5,6,7,8$ 
and a better fit $(\leq 1 \%)$ for $9 \leq n \leq 64$. 
(We used $n=65$ to set $c_0$, so it should not count.)
Let
\begin{equation}
f_n(x) = x^n \,P_{\rm fit}(x) = c_0 \,x^{n-2} \, e^{-a x^{1/3}} \,,
\end{equation}
so 
\begin{equation}
A_n = \int_{-x_0}^\infty f _n(x)\,dx 
\end{equation}
is our predicted moment from the above form. The maximum of 
the function $f_n(x)$  will be where $f'_n(x) = 0$, 
corresponding to 
\begin{equation}
x_{\rm max} = {\Bigg[\frac{3(n-2)}{a} \Bigg]}^3 \,.
\end{equation}
If $P_{\rm fit}(x)$ gives a good approximation for $A_n$, then 
it should give a good approximation to the exact $P(x)$ for $x \sim O(x_{\rm max})$.

For the electromagnetic energy density $a \approx 1$, so for 
$n=4$, $x_{\rm max} \approx 216$, and for $n=65$, 
$x_{\rm max} \approx 6751269$. Thus if $P_{\rm fit}$ gives 
reasonable fits to the moments for $4 \leq n \leq 65$, 
then it should be a fair
approximation to the exact distribution 
in the range, roughly, $10^2 \leq x \leq 10^7$, assuming uniqueness of the distribution.

\subsection{Approximate fits for $P(x)$ including the inner part}
\label{sec:appfits}

One can attempt to model the entire probability distribution, 
including the inner part, by experimenting with functions of the form:
\begin{equation}
P(x) = c_1 \,{(x_0+x)}^{-\alpha} \, {\rm exp}[-\beta {(x_0+x)}^\gamma]
 + c_0 \, {(\alpha_0+{(x_0+x)}^{2})}^{-1}\, \,{\rm exp}[-a {(x_0+x)}^{1/3}] \,.
\label{eq:pfit}
\end{equation}
A reason for using this form is that one need not bother with 
trying to match inner and outer parts of the function. 
Depending on the choices of the constants, one can possibly get 
the first term to dominate for small $x$, and the second for large
$x$. We use the values of $a, b_1$ from the tail fits and the
 values of $x_0$ from the quantum inequality bounds given earlier 
in Sec.~\ref{sec:estimates}.

The most interesting case is the distribution for $\rho_{EM}$, 
the electromagnetic energy density. For the values of the constants 
given in Table~\ref{table:fitconstants}, the fractional errors 
between the calculated and fitted moments in the $0^{\rm th}$ 
through $22^{\rm nd}$ moments are given in
Table~\ref{table:fracerrors}. 
Since the exact value of the first moment is 0, we list the 
fitted value separately as: $1^{\rm st}$ moment $= 0.0247001$.
 The errors in the fourth and fifth 
moments are somewhat large ($\sim 15\%$), but the errors tend to
 progressively decrease as we go to large $n$.  So this heuristic model 
 distribution gives 
 a reasonably good fit for the innermost part of the distribution and the tail, 
 but does somewhat poorly for the middle part of the distribution.

\begin{figure} 
\begin{center}
\includegraphics[width=9cm]{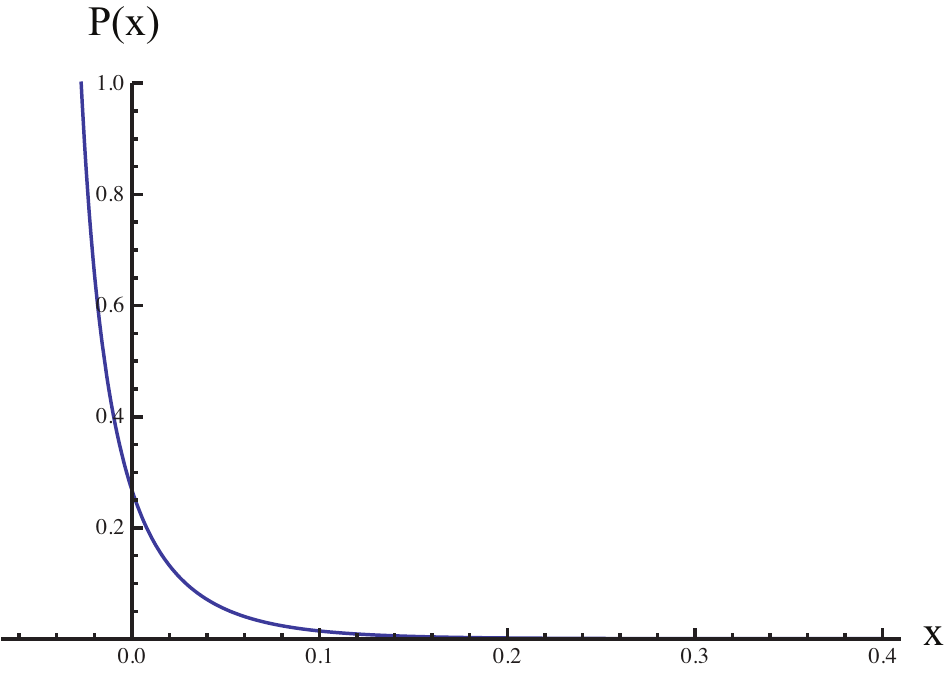}
\end{center}
\caption{The graph of $P(x)$ vs $x$ of our fit to the probability 
distribution function for $\rho_{EM}$, the electromagnetic energy density sampled
in time with a Lorentzian of width $\tau$. Here $x = 16 \pi^2 \tau^4 \, \rho_{EM}$.
The distribution has an 
integrable singularity at the conjectured optimal quantum inequality bound $x=- x_0= -0.0472$.} 
\label{fig:T00EMenergydensity} 
\end{figure}

The graph of $P(x)$ vs $x$ for this case is given in 
Fig.~\ref{fig:T00EMenergydensity}. It has a spike
(an integrable singularity) at the quantum
 inequality lower bound. However, our method may not be
sufficiently sensitive to conclude the existence of this 
singularity. It is possible that there are non-singular 
distributions which fit the first several moments as well as 
does our postulated form. Thus we cannot conclude 
 whether the 
actual distribution has a spike in it at the lower quantum 
inequality bound, as indicated in the plot. The distributions 
for $\rho_S$  and ${\dot \varphi}^2$ in two-dimensional 
spacetime, which are known exactly and uniquely, both have a 
spike at the quantum inequality lower bound, as does the 
distribution for $\varphi^2$ in four dimensions~\cite{FFR10}. 
In Tables~\ref{table:fitconstants} and \ref{table:fracerrors},
we list the fitting constants and fractional errors, respectively, 
for the $\rho_{EM}$ probability distribution. The values of the 
constants were obtained by calculating the moments from  
Eq.~(\ref{eq:pfit}) and using the  MATHEMATICA
{\bf Manipulate} command to adjust the values of the 
constants to get the smallest fractional errors between 
the fitted moments and the actual moments.

\begin{table}[htdp]
\caption{Fitting Constants for the Model Distribution for  $\rho_{EM}$
in Eq.~(\ref{eq:pfit}).}
\label{table:fitconstants}
\begin{center}
\begin{tabular}{|c|c|c|c|c|c|} \hline
 Constant  & $\rho_{EM}$    \\ \hline 
$a$ &  0.9630614156        \\ \hline
$c_0$ &  0.95539211         \\ \hline
$x_0$ & 0.0472       \\ \hline
$\alpha_0$ & 610          \\ \hline
$c_1$ & 0.028          \\ \hline
$\beta$ & 19.65         \\ \hline
$\gamma$ & 1.05         \\ \hline
$\alpha$ & 0.9999        \\ \hline
\end{tabular}
\end{center}
\end{table}

\begin{table}[htdp]
\caption{Table of Fractional Errors. Here the fractional error is $[a_n({\rm fit}) - a_n]/a_n$,
where the $a_n$ are given in Table~\ref{table:moments} , and the $a_n(\rm fit)$ are 
computed from Eq.~(\ref{eq:pfit}). For the $n=1$ case, the fractional error is not defined, 
since the first moment is 0. Fractional errors in succeeding moments beyond $n=5$ 
are progressively smaller. Although all moments through $n=65$ were used, we display
the fractional errors through $n=21$.} 
\label{table:fracerrors}
\begin{center}
\begin{tabular}{|c|c|c|c|c|c|} \hline
 $n$  &$\rho_{EM} $    \\ \hline 
0 &   0.00450644      \\ \hline
$1^{\rm st}$  moment & {\rm not applicable}      \\ \hline
2 &     -0.00661559      \\ \hline
3 &   -0.0770297         \\ \hline
4 &    -0.152164       \\ \hline
5 &    -0.150279        \\ \hline
6 &  -0.117773         \\ \hline
7 &      -0.0843077    \\ \hline
8 &   -0.0590582       \\ \hline
9 &  -0.0420107       \\ \hline
10 &    -0.0308225        \\ \hline
11 &  -0.0233756         \\ \hline
12 &  -0.0182526          \\ \hline
13 &   -0.0145945            \\ \hline
14 &     -0.0118911       \\ \hline
15 &   -0.00983456       \\ \hline
16 &  -0.0082327           \\ \hline
17 &    -0.00696063         \\ \hline
18 &   -0.00593416          \\ \hline
19 &    -0.00509465          \\ \hline
20 &    -0.00440012          \\ \hline
21 & -0.00381978              \\ \hline
\end{tabular}
\end{center}
\end{table}

\section{Bounds on the cumulative distribution function}\label{sect:CDF}

As already mentioned, it is possible that the moment problem is indeterminate 
and that there are many probability distributions with these moments. Here, we
show that no such distribution can have a tail decreasing much more slowly than
that studied above. Our tool for this purpose is a simple variant of Chebyshev's
inequality: if $X$ is any random variable taking values in $[-x_0,\infty)$, 
with moments $a_n$, then the probability $\text{Prob}(X\ge \lambda)$ that $X$ exceeds any given $\lambda$ is bounded by 
\begin{equation}
\text{Prob}(X\ge \lambda) \le  \frac{a_n +  \text{Prob}(X<0) x_0^n}{\lambda^n}
\end{equation}
for all $n$. To prove this, let $d\mu(x)$ be the probability measure of $X$ and then
compute
\begin{equation}
\lambda^n \text{Prob}(X\ge \lambda) = \lambda^n \int_\lambda^\infty d\mu(x) \le 
\int_\lambda^\infty x^n d\mu(x) \le 
\int_0^\infty x^n d\mu(x) = a_n - \int_{-x_0}^0 x^n d\mu(x)
\le a_n + \text{Prob}(X<0) x_0^n .
\end{equation}
(The term $\text{Prob}(X<0) x_0^n$ is only needed for odd $n$, in fact. We have also written 
$d\mu(x)$, rather than $P(x)dx$ for the probability measure to emphasise that we are not assuming a 
continuous probability density function.) In our case, we know
that $x_0<x_0(FE)<1$, so we have 
\begin{equation}
\text{Prob}(X\ge \lambda) \le \inf_{n\in\mathbb{N}} \frac{a_n+1}{\lambda^n}.
\end{equation}
Now, for moments growing as $a_n \sim C D^n (3n-4)!$,  the ratio of successive terms in the
infimum is 
\begin{equation}
\frac{a_{n+1}+1}{\lambda(a_n+1)} \sim D\frac{(3n-1)(3n-2)(3n-3)}{\lambda},
\end{equation}
so, for each fixed $\lambda$, the sequence will decrease until the term where 
$n\sim \frac{1}{3} (\lambda/D)^{1/3}$ and will increase thereafter. This gives an asymptotic bound on 
the tail probability
\begin{equation}
\text{Prob}(X\ge \lambda) \lesssim C\left(\frac{D}{\lambda}\right)^{\frac{1}{3}(\lambda/D)^{1/3}}
\Gamma\left((\lambda/D)^{1/3}-3\right)\sim
\sqrt{2\pi} C \left(\frac{D}{\lambda}\right)^{7/6} e^{-(\lambda/D)^{1/3}}.
\end{equation}
as $\lambda\to\infty$. 

This gives an upper bound on the tail probability distribution, which is not much more slowly
decaying than that for our fitted tail, for which the tail probability would be decaying like
$C \left(\frac{D}{\lambda}\right)^{4/3} e^{-(\lambda/D)^{1/3}}$. The following 
discussion sketches how information on the lower bound can be obtained; this could be developed 
into a rigorous discussion (and probably sharpened) with further work. In fact, we do not seek 
a strict lower bound on the tail probability, but rather aim to show that it must be very often of the 
order of the fitted tail or higher. 

Let $Q(x)=\text{Prob}(X\ge x)$. Then we have, for any $\Lambda>\lambda>x_0$,
\begin{align}
a_n &\le \lambda^n\text{Prob}(X<\lambda) + 
\int_\lambda^\infty x^n d\mu(x)  \\ &\le
\lambda^n\text{Prob}(X<\lambda) + Q(\lambda)\lambda^n +n \int_\lambda^\infty 
Q(x) x^{n-1}\,dx  \\
&\le  \lambda^n +n \int_\lambda^\Lambda 
Q(x) x^{n-1}\,dx  + \sqrt{2\pi} C D^{n-1} n \int_\Lambda^\infty
\left(\frac{x}{D}\right)^{n-13/6} e^{-(x/D)^{1/3}}\,dx \\
&\le \lambda^n +n \int_\lambda^\Lambda 
Q(x) x^{n-1}\,dx  + 3\sqrt{2\pi} C D^{n} n \Gamma( 3n-7/2, (\Lambda/D)^{1/3})
\label{eq:an_upper}
\end{align}
in which we have integrated by parts in the second line and used the fact that $Q(\lambda)=
1-\text{Prob}(X<\lambda)$, as well as the upper bound found above; $\Gamma(N,z)$ is
the upper incomplete $\Gamma$-function. We can now
make $n$-dependent choices of $\lambda$ and $\Lambda$ so that the first and third terms are 
negligible in comparison with $a_n$ for large enough $n$.  For example, 
$\Lambda = (4n)^{3}D$ and $\lambda = n^3 D$ will do: it is a simple
application of Stirling's formula to see that 
$\lambda^n/(D^n(3n-4)!) \sim \text{const}\times n^{7/2} (e/3)^{3n} \to 0$; 
for the upper end we first estimate
$\Gamma(3n-7/2,4n) \sim 4 (4n)^{3n-9/2}e^{-4n}$ 
using Laplace's method (see~\cite{deBruijn}, section 4.3) \footnote{We have
$\Gamma(3n-7/2,4n)=\int_{4n}^\infty y^{3n-9/2}e^{-y}dy=
(4n)^{3n-7/2}e^{-4n}\int_0^\infty 
(1+u)^{-9/2} e^{n (-4u + 3\log(1+u))}du$, under the change of variable
$u=y/(4n)-1$.  The integral is simply estimated as $1/n$ by
the method of Laplace~\cite{deBruijn}.} which gives
\begin{equation}
n\frac{\Gamma(3n-7/2,4n)}{(3n-4)!} 
\sim \frac{1}{\sqrt{2\pi}}\left(\frac{3}{4}\right)^{7/2}
\left(\frac{64}{27e}\right)^n \to 0.
\end{equation}

With these choices of $\lambda$ and $\Lambda$ in force, we set  
$F(x) = xQ(x)e^{(x/D)^{1/3}}$, 
whereupon we have
\begin{equation}
n \int_\lambda^\Lambda F(x) x^{n-2}e^{-(x/D)^{1/3}}\,dx \gtrsim CD^n (3n-4)!
\end{equation}
from \eqref{eq:an_upper}. Now let $S$ be the subset of $x\in [\lambda,\Lambda]$ for which 
$F(x)\ge \frac{1}{2} CD (D/x)^{1/3}$. We bound $F$ from above by $\sqrt{2\pi}CD (D/x)^{1/6}$ on $S$,
and by $\frac{1}{2} CD (D/x)^{1/3}$ on the complement $S^c$ of $S$ in $[\lambda,\Lambda]$,
to give
\begin{equation}
\int_\lambda^\Lambda F(x) x^{n-2}e^{-(x/D)^{1/3}}\,dx 
\le  \sqrt{2\pi}CD^{7/6}\int_{S} x^{n-13/6}e^{-(x/D)^{1/3}}\,dx  +  
\frac{CD^{4/3}}{2} 
\int_{S^c} x^{n-7/3} e^{-(x/D)^{1/3}}\,dx.
\end{equation}
Now the first integral on the right-hand side can be bounded from above by the supremum of 
the integrand
multiplied by the Lebesgue measure $|S|$ of $S$, while the second is bounded by the integral 
over
all $[0,\infty)$. The supremum mentioned occurs for $x=(3n-13/2)^3D$, and we find
\begin{equation}
CD^n (3n-4)!
\lesssim |S| \sqrt{2\pi} CD^{n-1}n (3n-13/2)^{3n-13/2}e^{-(3n-13/2)}  + \frac{1}{2}CD^n 3n
\Gamma(3n-4).
\end{equation}
Rearranging and using Stirling's formula, this requires
\begin{equation}
|S| \gtrsim \frac{(3n-4)!e^{3n-13/2}D}{\sqrt{8\pi} n (3n-13/2)^{3n-13/2}}
\sim
\frac{27}{2} D n^2 .
\end{equation}

Summarizing, we have shown that in the interval $[n^3 D, 4n^3 D]$, for $n$ sufficiently large, we have
\begin{equation}
\text{Prob}(X\ge x) \ge  \frac{1}{2}C \left(\frac{D}{x}\right)^{4/3}  e^{-(x/D)^{1/3}}
\end{equation}
on a set with measure at least $\frac{27}{2} D n^2$.  It seems likely that this is a
substantial underestimate of the measure of $S$, as some of the estimates used in the last part of the 
argument are rather weak. 

Thus the broad behavior of the tail of the probability distribution is determined by the moments, even if 
the exact probability distribution is not uniquely determined. In the applications we give below, it is 
only the broad behavior that is required.

\section{Possible Applications for the Tail}
\label{sec:apps}

\subsection{Black Hole Nucleation}
\label{sec: bhn}
The fact that the energy density probability distribution has 
a long positive tail implies a finite probability for the 
nucleation of black holes out of the Minkowski vacuum via 
large, though infrequent positive fluctuations. 
This probability cannot be too large, of course, or it will 
conflict with observation. Let us sample a spacetime region 
(a cell) over a size $\ell \approx \tau$, where $\tau$ equals the 
sampling time. For an energy density $\rho$, which is roughly
constant in space, the associated mass will be
 $M \approx \rho {\ell}^3$. 
This can be a black hole if $GM \approx \ell$, or 
${\ell_p}^2 M \approx \ell$, in units where $\hbar=c=1$ 
and $\ell_p$ is the Planck length, which implies 
$\rho \approx 1/({\ell_p}^2 {\ell}^2)$. Here we chose  $\tau \approx \ell$,
so that the sampling time is approximately the light travel time across
the black hole.

Note that we should really use the probability distribution for energy
density sampled over a spacetime volume, with the spatial and
temporal dimensions approximately equal. For the purpose of an
order of magnitude estimate, we assume that the probability
distribution for sampling in time alone will yield roughly similar
results. 

Let our observation volume be $V$ and our total observation time 
be $T$. The number of cells in this spacetime volume is 
$N=VT/{\ell}^4$. Because black hole nucleation will be a rare event,
we  assume that different nucleation events will be widely
separated and  uncorrelated. The number of black holes, $n$, nucleated 
in this spacetime volume, $VT$ is then $n \approx N P_n$, where 
$P_n$ is the probability of a black hole nucleation in our 
sampled spacetime volume ${\ell}^4$. Let us estimate that
\begin{equation}
P_n \approx \int_x^{2 x} P(y)\, dy \,.   \label{eq:Pn}
\end{equation}
where
\begin{equation}
x=16\pi^2 \tau^4 \rho = 16 \pi^2 \frac{\ell^2}{{\ell_p}^2}=
16 \pi^2 {\Bigg(\frac{M}{m_p}\Bigg)}^2 \,,
\end{equation}
and $m_p$ is the Planck mass. Here $P_n$ is the probability of nucleating
a black hole in the range between $x$ and $2x$. However, in the limit
of large $x$, $P_n$ will be independent of the exact upper limit in 
Eq.~(\ref{eq:Pn}).  Let the probability 
distribution have a tail of the form given by Eq.~(\ref{eq: pb2}). Then
\begin{equation}
P_n \approx c_0 \int_x^{2x} y^{-2} \, e^{-a y^{1/3}}\, dy =
3 c_0 a^3 \int_{u_1}^{u_2} u^{-4} e^{-u} du = 3\, c_0\, a^3 [\Gamma(-3,u_1)
-\Gamma(-3,u_2)] \,.
\end{equation}
Here $u = a y^{1/3}$, $u_1 = a x^{1/3}$,  $u_2 = 2^{1/3}\, u_1$, and 
$\Gamma(-3,u)$ is an incomplete gamma function. This function has the
asymptotic form
\begin{equation}
\Gamma(-3, u) \approx u^{-4}\, e^{-u}
\end{equation}
for $u \gg1$. From this form, we see that the contribution from the lower integration
limit dominates, and we have
\begin{equation}
P_n \approx \frac{3 \, c_0}{a}\, x^{-\frac{4}{3}} \, e^{-a x^{1/3}}  \label{eq:P_n}
\end{equation}
for large $x$.

Thus we have for the mean number of
nucleated black holes
\begin{equation}
n = \frac{VT}{\ell^4} P_n = \frac{VT}{{\ell_p}^8 M^4} P_n \,,
\end{equation}
or, using Eq.~(\ref{eq:P_n}), 
\begin{equation}
n \approx \frac{3 c_0}{a} \, (16 \pi^2)^{-4/3}
\left( \frac{VT}{{\ell_p}^4}\right) \,{\Bigg( \frac{m_p}{M}\Bigg)}^{20/3} 
{\rm exp}[{-a_0 {(M/m_p)}^{2/3}}] \,,
\end{equation}
where $a_0 = {(16 \pi^2)}^{1/3} a$. For the energy density 
of the EM field, $c_0 \approx0.955,\,a \approx 0.963$, 
so $a_0 \approx 5.2$. Therefore for this case we have
\begin{equation}
n \approx 10^{-2} \, 
\Bigg( \frac{VT}{{\ell_p}^4}\Bigg) \,{\Bigg( \frac{m_p}{M}\Bigg)}^{20/3}
{\rm exp}[{-5.2 {(M/m_p)}^{2/3}}] \,.
\label{eq: nEM}
\end{equation}

To estimate the probability of black hole nucleation, let us first 
choose $V=1 {\rm cm}^3$, $T= 1$ sec, and $n=1$, which gives 
$VT/{\ell_p}^4 \approx {(10^{33})}^3 \,10^{43} \approx 10^{142}$. 
We want the probability of one black hole forming in one cubic 
centimeter of space over an observation time of one second. We can 
use Eq.~(\ref{eq: nEM}) to determine the resulting mass of the black 
hole, which turns out to be $M \approx 400 \, m_p$. Let us now
consider our observation volume and time to be the size and age 
of the universe, which gives $VT/{\ell_p}^4 \approx
{(10^{28}/10^{-33})}^4 \approx 10^{244}$. 
Taking $n=1$ again, and using Eq.~(\ref{eq: nEM}), yields $M \approx
990 \, m_p$. 
Therefore, if we observe a volume the size of the universe for 
a time equal to the age of the universe, we are likely to see 
the nucleation of only about one $10^3 m_p$ black hole from the vacuum.

Thus nucleation of black holes of mass $\sim 10^2 \, m_p$ is common, 
but $10^3 m_p$ black holes are very rare. Why do we not notice these 
$400 \, m_p \approx 10^{-2}$ g black holes? Presumably they must 
appear for a very short time and be surrounded by negative energy 
which quickly destroys them.

\subsection{Boltzmann brains}
\label{sec: bbs}
Recently, the ``Boltzmann brain'' problem has become the 
subject of increasing interest in cosmology~\cite{Bbrains,Brains_olum}.
This is the possibility that conscious entities, which may or may not
resemble biological brains, might spontaneously nucleate and
exist for a finite time. Anthropic reasoning requires a count
of observers, as the anthropic prediction for the value of an
observable is the value most likely to be found by a typical
observer. If the typical observer is a Boltzmann brain in
intergalactic space, and not an observer on an earthlike planet,
this would greatly alter anthropic predictions.  As a somewhat more
speculative application,
we consider what the tails of our probability distributions 
have to say about the probability of nucleating Boltzmann 
brains in four-dimensional Minkowski spacetime. This 
calculation is similar to the one above for the nucleation of black holes.

Consider a spatial region of size $\ell$, a timescale $\tau$, 
and a mass $M$, so that the mean energy density is $\rho \approx
M/{\ell}^3$.
 We want to use the tail of the EM energy density probability
 distribution to estimate the probability of mass $M$ appearing 
in this specific region in a particular interval $\tau$. Our 
sampled energy density is $x=16 \pi^2 {\tau}^4 \,\rho 
\approx {\tau}^4 \, M/{\ell}^3$. So we have that 
\begin{equation}
P(x) \propto e^{-a x^{1/3}} \approx e^{- x^{1/3}} \approx 
{\rm exp}{\Bigg(-{\tau}^{\frac{4}{3}} \frac{M^{1/3}}{\ell}\Bigg)} \,
\end{equation}
where we have ignored the prefactor and used $a\approx 1$. 
The prefactor would contain information about the fraction of 
mass $M$'s that could think. Even if very small, this factor is
likely to pale in importance compared to the exponential factor 
derived below. Let 
$M=1\, {\rm kg} \approx 10^{41} \,{\rm cm}^{-1}$, $\ell = 10$ cm, 
and $\tau = 0.3 \, {\rm sec} \approx 10^{10}$ cm. These values give
\begin{equation}
{\tau}^{\frac{4}{3}} \frac{M^{1/3}}{\ell} \approx 10^{26} \,,
\end{equation}
so
\begin{equation}
P \approx e^{-10^{26}} \,.
\end{equation}
This is much larger than the ${\exp}(-10^{50})$ estimate of 
Page~\cite{Page}, who assumes that $P \propto e^{-I}$, where 
$I=M t=$ action. So our energy density probability distribution 
makes the Boltzmann brain problem worse. Although the probability
per unit volume for the nucleation of a Boltzmann brain may seem
exceedingly low, the available volume could make them more numerous 
than other observers. Note that in this case, the energy density has been 
averaged over a spacetime region which is much larger in the time direction
than in the spatial directions, $\tau \gg \ell$. Hence the probability
distribution for the energy density averaged in time alone should be a good 
approximation here.

\section{Discussion}
\label{sec:diss}

\subsection{Uniqueness Issues}
\label{sec:unique}

As was noted in Sec.~\ref{sec:gen}, the moments which we calculate
for ${:}\dot{\varphi}^2{:}$ and related operators satisfy neither the Hamburger condition,  
Eq.~(\ref{eq:Hamburger}), nor the Stieltjes condition, Eq.~(\ref{eq:Stieltjes}) for uniqueness. 
Thus none of
our results for $P(x)$ are rigorously guaranteed to be unique. However, there
are some observations which are relevant here. First, these are sufficient, but
not necessary, conditions for uniqueness. There is a necessary and sufficient
condition~\cite{Simon}, but this condition requires detailed knowledge of all 
moments and does not seem to be testable in our problem. Second, rapid growth
of moments does not automatically mean non-uniqueness. There are examples
of sets of moments which grow at arbitrary rates, but nonetheless are associated
with unique probability distributions.

On the other hand, if the probability distribution is continuous, with probability density function $p(x)$
on $[-x_0,\infty)$, and 
\begin{equation}
\int_{-x_0}^\infty \frac{\log(p(x))\,dx }{\sqrt{x+x_0}(1 + x)} > -\infty
\end{equation}
then the Stieltjes problem is indeterminate for the moments of $p$ (assuming they all exist,
and that $x_0<1$ for convenience); there
is more than one probability distribution supported in $[-x_0,\infty)$ with the same moments. This is 
a theorem of Krein (modified slightly to our setting) see, e.g., Theorem~5.1 in Ref.~\cite{Berg}.
In particular, 
this would show that any distribution whose tail was exactly equal to 
$P_{\rm fit}(x) = c_0 x^{-2}e^{-ax^{1/3}}$ for large enough $x$ had indeterminate moments in the 
above sense. On the other hand, if $p(x)$ were to oscillate around $P_{\rm fit}(x)$, but sometimes 
taking much smaller values than $P_{\rm fit}$, then the logarithm will take large negative values; 
such behavior could lead the integral to diverge 
and allow the moment problem to be determinate. 

To illustrate how delicate the uniqueness issue can be, we note that the probability distribution
$P(x) = \frac{1}{6}\theta(x)e^{-x^{1/3}}$, has moments $a_n = \frac{1}{2}(3n+2)!$, 
that are indeterminate in the Stieltjes sense on $[0,\infty)$ by Krein's theorem. 
However, mild modifications of these moments yield determinate problems. For example, 
by Cor.~4.21 in Ref.~\cite{Simon}, there exists a constant $c$ so that the set of moments 
$\tilde{a}_0=1$, $\tilde{a}_{n}=c (3n-1)!$ is a {\em determinate} problem, corresponding to a purely 
discrete probability distribution. 

Overall, we are not able to resolve the question of determinacy, although on balance our expectation
is that the problem is indeed indeterminate. Certainly we have not been able to find any positive
evidence to suggest that the moments are determinate. Nonetheless, certain features of the
probability distribution can be ascertained. We have shown in Appendix~A that our moments 
grow as a power times $(3n-4)!$.
This rate of growth seems to be just what is needed to produce distributions
with tails falling as in  Eq.~(\ref{eq: pb2}), that is, proportional to $x^{-2}e^{-a x^{1/3}}$.
We have argued that any probability distribution arising from our moments will have a broadly 
similar tail. This asymptotic behavior is all
that is needed for many applications of our distributions, such as those discussed
in Sect.~\ref{sec:apps}.  

It is also worth noting that the conclusion that the probability distribution has a 
lower bound is independent of any concerns about uniqueness,
because this follows from existing quantum inequality bounds. Our actual estimates
of the lower bounds, given in Sect.~\ref{sec:estimates}, are not strictly independent
of the uniqueness issue, but only use a finite number of the moments. Thus the
numerical answers obtained only depend upon the values of these moments.

\subsection{Summary}
\label{sec:sum}

In this paper we have explored possible probability distributions for averaged quadratic
operators in the four-dimensional  Minkowski vacuum state. 
We use averaging with a Lorentzian function of
time, and investigate the distributions for $\dot{\varphi}^2$, where $\varphi$ is a massless
scalar field, for $\rho_S$, the associated scalar field energy density, for $E^2$, the squared
electric field, and for $\rho_{EM}$. In all cases, we infer that the distributions have some
features in common with our previous results~\cite{FFR10} for a conformal field in
 two dimensions and for $\varphi^2$ in four dimensions. Specifically, there is a
 lower bound on the distribution, which coincides with the optimal quantum inequality
 bound on the associated expectation value in an arbitrary quantum state. Furthermore,
 there is no upper bound on the distributions, so arbitrarily large positive quantum
 fluctuations are possible. 

We have outlined a procedure that, in principle, allows the calculation
of an arbitrary number of moments of a given distribution. In practice,
this procedure can be carried at least as far as the $65^{th}$ moment,
which is sufficient to allow reasonable numerical estimates of both
the lower bounds, and of the asymptotic tail for large argument. These
are not guaranteed to be unique, but as was argued in the previous
subsection, they may be useful.

If we accept the forms of the tails which we find, then several
physically
interesting applications follow, including the nucleation rates for
black holes and Boltzmann brains. It should also be possible to apply
these results to the study of the small scale structure of four
dimensional spacetime, along the lines studied in two dimensions in 
Ref.~\cite{CMP11}.  It may also be possible to learn more about the
non-Gaussian density and gravity wave perturbations in inflationary
cosmology, which were studied in Refs.~\cite{WKF07,FMNWW10,WHFN11}. 
Another implication of our form for the tail is that vacuum fluctuations will
dominate thermal fluctuations at high energies. The   Boltzmann 
distribution falls exponentially with energy, but vacuum energy density
fluctuations fall more slowly and hence eventually dominate.

There is clearly room for further work on the topic of this paper. One obvious
problem is to determine whether or not the moment problems we have studied
are determinate: if so, one would like to know the detailed form of the corresponding
probability distributions; if not, one would like to know how much information
may be extracted from the moments, nonetheless, along the lines of the arguments
in Sect.~\ref{sect:CDF}. In addition, our results have now trapped the sharp quantum inequality 
bounds for various operators between the lower
bounds given by the methods of Ref.~\cite{FewsterEveson98} 
and the bounds obtained in Sect.~\ref{sec:estimates}, which are an order of magnitude smaller. 
If the moment problem is determinate, the latter
bounds will coincide with the sharp bound; otherwise it would be interesting to
determine what the sharp bound actually is. Recall that here we deal only with Lorentzian sampling
and only in the time direction. It will also be of interest to investigate more general
sampling functions, and the effects of sampling in space as well as time.

\begin{acknowledgments}
This work was supported in part by the National
Science Foundation under Grants PHY-0855360 and PHY-0968805.
\end{acknowledgments}

\appendix
\section{Computation of the moments}\label{appx:comp}

We describe how the moments of smeared Wick squares may be computed for
a general derivative $\phi$ of the massless field $\varphi$ in four dimensions, writing $p$ for one 
more than twice the number of derivatives,
so $p=1$ for ${:}\varphi^2{:}$ and $p=3$ for  ${:}\dot{\varphi}^2{:}$. 
Thus the two-point function for $\phi$, restricted to the time axis, is given by
\begin{equation}
\langle \phi(t)\phi(t')\rangle = \frac{1}{4\pi^2}\int_0^\infty d\omega\,\omega^p e^{-i\omega
(t-t'-i\epsilon)}\,.
\end{equation}
With smearing along the time axis against smearing function $f$, the rules for computing the 
contribution to the $n$'th connected moment of a given connected graph on $n$ vertices may be 
stated in Fourier space as follows. For each line, 
 the form of the two-point function
entails that there is a momentum integral over the positive half-line 
and a factor of the $p$'th power of the momentum; for each vertex there is a factor of 
$\hat{f}(\omega_j+\omega_k)$ if the vertex is the source of the lines carrying momenta 
$\omega_j$ and $\omega_k$, a factor of $\hat{f}(\omega_j-\omega_k)$ if the vertex is the source 
(resp., target) of the line carrying
momentum $\omega_j$ (resp., $\omega_k$), or a factor
of $\hat{f}(-\omega_j-\omega_k)$ if the vertex is the target of the lines carrying momenta $\omega_j$ 
and $\omega_k$; there is an overall factor of $(4\pi^2)^{-n}$ and a combinatorial factor that is $2^n$ 
for $n\ge 3$ and $2$ for $n=2$. Here $\hat{f}$ is the Fourier transform, defined with the convention 
\begin{equation}
\hat{f}(\omega) = \int_{-\infty}^\infty dt f(t)e^{i\omega t} \,.
\end{equation}
An important point is that if, as for the Lorentzian, $\hat{f}$ is real and positive, then every graph 
contributes positively to the moment. Thus any individual graph on $n$-vertices gives a lower bound 
on the $n$'th connected moment. If one wishes
to compute the dimensionless moments, defined in the text so that 
$a_n=(4\pi \tau^{(p+1)/2})^{2n}\mu_n$, the overall factor $2^n (4\pi^2)^{-n}$ is replaced by 
$8^n \tau^{-(p+1)}$ for $n\ge 3$ (or $32$ in the case $n=2$).

In the particular case of the Lorentzian~\eqref{eq:lorentzian}, we have 
$\hat{f}(\omega) = {\rm e}^{-|\omega|\tau}$, and a simplification of the computation rules:
a vertex met by lines carrying momenta $\omega_j$ and $\omega_k$  contributes
 ${\rm e}^{-|\omega_j-\omega_k|\tau}$ if it is a target for one and a source for the other, 
or $e^{-(\omega_j+\omega_k)\tau}$ otherwise. This means that the overall
integral over $\omega_1,\ldots,\omega_n$ factorizes at each vertex that is
either a double source or a double target. 

Recall that the graphs involved are drawn on $n$ vertices $x_1,\ldots,x_n$,
placed in increasing order from left to right. Each vertex is met by two distinct lines,
and each line is directed to the right. In particular,  $x_1$ is the source of
both lines connected to it. We may represent such a graph by 
a permutation $\sigma$ of the set $\{1,\ldots,n\}$ of integers, subject
to the conditions that $\sigma(1)=1$ and $\sigma(2)<\sigma(n)$. 
To reconstruct a graph from a permutation,
draw lines from $x_1 = x_{\sigma(1)}$ to $x_{\sigma(2)}$, from
$x_{\sigma(2)}$ to $x_{\sigma(3)}$, and so on, finishing with an line
from $x_{\sigma(n)}$ to $x_1$. [For this reason, it is convenient to adopt a convention that 
$\sigma(n+1)=1$.] Then place rightwards-pointing arrows on each line. On the other hand, to encode 
a graph as a permutation, start at $x_1$
and follow the shorter of the two lines to the vertex it meets [i.e., of the two
vertices joined to $x_1$, choose the one with the smaller label], and continue
along the other line meeting that vertex. At each subsequent vertex, 
continue along the line not previously traversed, eventually returning to 
$x_1$. Then $\sigma(k)$ is defined to be the $k$'th vertex met on this trip.

A {\em run} of $\sigma$ is a set of consecutive integers in $\{1,\ldots,n+1\}$,
say $p,p+1,\ldots q$, such that  $\sigma(p),\sigma(p+1),\ldots, \sigma(q)$ is a monotone sequence, 
either ascending or descending, and so that no superset of consecutive integers has the same 
property. The length of the run is defined to be $q-p$. Every permutation used to label our graphs 
corresponds to an  even number of runs, alternating between ascending and descending, whose 
lengths sum to $n$, and with consecutive runs sharing a common endpoint. 
For example, the permutation $14536782$ (i.e., $\sigma(1)=1$, $\sigma(2)=4$,
$\sigma(3)=5$,...) has runs $1,2,3$; $3,4$; $4,5,6,7$ and $7,8,9$, of lengths $2,1,3,2$; representing 
each run by its image under the permutation, these runs are more transparently written as 
$145$, $53$, $3678$, $821$.

The contribution to the $n$'th connected moment arising from the graph 
corresponding to any given permutation is easily seen to factorize into terms corresponding to the 
runs, whose values depend only on the length of the run: in our example, the graph contributes 
$8^8K_2 K_1 K_3 K_2 = 8^8 K_1 K_2^2 K_3$
to the dimensionless connected moment $C_8$,
where the $K_j$ correspond to the special case $K_j=K_j^{(0)}$ of the family of integrals
\begin{equation}\label{eq:Knr_def}
K_n^{(r)} =  \frac{2^r}{r!}
\int_{(\RR^+)^{\times n}} dk_1\,dk_2\,\cdots dk_{n}
k_1^{p+r} (k_2\cdots k_{n})^p e^{-k_1}
e^{-\sum_{i=1}^{n-1} |k_{i+1}-k_i|}e^{-k_{n}}.
\end{equation}
[Here, $k_i=\omega_i \tau$ are dimensionless versions of the momenta previously used.]

These considerations reduce the computation of the $n$'th connected
moment to two problems: the computation of the $K_j$ and the enumeration
of all permutations in the class considered with a given run structure. To
address the first of these, we note the easily proved identity
\begin{equation}\label{eq:observation}
\int_0^\infty dk\, k^q e^{-k}e^{-|k-\kappa|} = \frac{q!e^{-\kappa}}{2^{q+1}}
\sum_{r=0}^{q+1} \frac{(2\kappa)^r}{r!},
\end{equation}
of which the standard integral
\begin{equation}\label{eq:special}
\int_0^\infty dk\,k^p e^{-2k} = \frac{p!}{2^{p+1}}
\end{equation}
is the $\kappa=0$ special case, and which entails the recurrence
relation
\begin{equation}\label{eq:recurrence}
K_n^{(r)}= \frac{p!}{2^{p+1}} \binom{p+r}{p}\sum_{r'=0}^{p+r+1} K^{(r')}_{n-1}.
\end{equation}
As $K_1^{(r)}= 2^{-(p+1)}p! \binom{p+r}{p}$, it follows that $K^{(r)}_n$ is given by
\begin{equation}
K_n^{(r)}= \left(\frac{p!}{2^{p+1}}\right)^n \binom{p+r}{p}
\sum_{r_{n-1}=0}^{p+1+r}\sum_{r_{n-2}=0}^{p+1+r_{n-1}}\cdots \sum_{r_{1}=0}^{p+1+r_2}
\prod_{k=1}^{n-1}\binom{p+r_k}{p}      \label{eq:Knr}
\end{equation}
for any integers $n\ge 1$ and $r\ge 0$. Although we have not found a
closed form expression for the $K_n^{(r)}$, the above expressions
allow for them to be computed efficiently. 

To the best of our knowledge, the problem of enumerating permutations of the class we study in terms 
of their run structure does not appear to have been solved in the literature on enumerative 
combinatorics, although related problems have been studied for over a century. A closed form answer 
seems out of reach, but generating
function techniques allow one to build up a solution for each $n$ in a recursive
way. The details will be reported elsewhere~\cite{Few_inprep}, but the overall
result is the following: for each $n$,  let $\mathcal{K}_n$  be a polynomial
in the variables $K_1,\ldots, K_{n-1}$, with $\mathcal{K}_2=\frac{1}{2} K_1^2$
and subject to the recurrence relation
\begin{equation}
\mathcal{K}_{n} = 
\sum_i K_{i+1}\frac{\partial  \mathcal{K}_{n-1}}{\partial K_i} + 
\sum_{i,j} K_1 K_i K_j \frac{\partial \mathcal{K}_{n-1}}{\partial K_{i+j}}.
\end{equation}
Then, for $n\ge 3$, the coefficient of $K_1^{m_1}\cdots K_{n-1}^{m_{n-1}}$
in $\mathcal{K}_n$  is precisely the number of permutations $\sigma$ of $\{1,\ldots,n\}$ with  $m_\ell$ 
runs of length $\ell$ ($1\le\ell\le n-1$), subject to the side conditions 
$\sigma(1)=1$, $\sigma(2)<\sigma(n)$. In the case $n=2$, we find
half of the number of such permutations. 

The generating function is extremely convenient, because it already incorporates
the sum over all possible connected graphs. Putting this together with the other considerations
above, the $n$'th dimensionless connected moment is given by
$C_n = 8^n \mathcal{K}_n$, for any $n\ge 2$, where the variables $K_j$
are given the values defined above by \eqref{eq:Knr_def} (recalling that
$K_j=K_j^{(0)}$). For example, we find the explicit formulae:
\begin{align}
C_2 & = 32 K_1^2 \\
C_3 & = 8^3 K_2 K_1 \\
C_4 & =8^4 \left(K_3K_{1}+ K_{2}^{2}+K_{1}^{4} \right) \\
C_5 & =8^5 \left(K_{4}K_{1} +  3 K_{3}K_{2}+8K_{2}K_1^{3}\right) \\
C_6 &=8^6 \left(K_5 K_1+3 K_3^{2}+4K_4 K_2+13 K_3K_1 ^{3}+31 K_2^{2}K_1^{2}+ 8 
K_{1}^{6}\right) \\
C_{7}&= 8^7\left(K_6K_1+10K_4K_3+5K_5K_2+19K_4K_1^{3}+66K_2^{3}K_1
+123K_3K_1^{2}K_2 + 136 K_2K_1^{5}\right)
\end{align}
which can be used to provide the first few connected moments for ${:}\varphi^2{:}$ in the case $p=1$ 
or ${:}\dot{\varphi}^2{:}$ in the case $p=3$. One may check that the coefficients inside each 
parenthesis sum to $(n-1)!/2$, the total number of connected graphs involved in the $n$'th moment.

\section{Asymptotics of the moments}\label{appx:asymp}

In this appendix we give asymptotic estimates for the $n$'th moments
of the Lorentzian smearing of the Wick square of the $\frac{1}{2}(p-1)$'th derivative of $\varphi$ as 
$n$ becomes large. We rigorously establish a lower bound
and also give an upper bound, for which our reasoning is not completely rigorous,
but which appears to be satisfied on the grounds of numerical evidence.  
The basic observation is that the dominant contribution to $C_n$ [and hence
the full dimensionless moment $a_n$] is $8^nK_{n-1} K_1 $; this is certainly a lower bound (as all 
terms are positive) and numerical evidence suggests that it gives the correct answer modulo a 
fractional error
of order $n^{-2}$. Thus lower bounds on the $K_{j}$ will give rigorous
lower bounds on $C_n$, while upper bounds give an upper bound on the 
$C_n$ that seems reasonably secure, albeit not fully rigorous. In terms
of permutations and graphs, the dominant contribution arises from the 
identity permutation, and thus the graph on $n$ vertices that has lines from $x_k$ to $x_{k+1}$ 
for each $k=1,\ldots,n-1$ and an line from $x_1$ to $x_n$. The graphs
in Fig.~\ref{fig:graph}  represent the case $n=2$ and $n=3$. 

We begin with the lower bound, which is
\begin{equation}\label{eq:lower}
K^{(r)}_n \ge \binom{n(p+1)-1+r}{n(p+1)-1}  L^{(r)}_n,
\end{equation}
where
\begin{equation}
L^{(r)}_n =  \frac{(n(p+1))!}{n! (2^{p+1}(p+1))^n} \prod_{k=1}^{n-1}\frac{r+n(p+1)}{r+k(p+1)},  
\end{equation}
in which the product over $k$ is taken to be equal to unity in the case $n=1$.
The bound~\eqref{eq:lower} is proved by induction, noting that the it is true 
(indeed, an equality) for $n=1$. 
Supposing that it holds for some $n\ge 1$, we use the recurrence relation Eq.~\eqref{eq:recurrence} 
to show that 
\begin{align}
K^{(r)}_{n+1} &\ge \frac{p!}{2^{p+1}}  \binom{p+r}{p}
\binom{(n+1)(p+1)+r}{n(p+1)} L_n^{(p+1+r)} \notag \\
&=\frac{p! }{2^{p+1}} 
\frac{r+(n+1)(p+1)}{r+p+1} \binom{(n+1)(p+1)-1}{p} \binom{(n+1)(p+1)-1+r}{(n+1)(p+1)-1}L_n^{(p
+1+r)},
\end{align}
where, in the first line, we have used the fact that the constants $L^{(r)}_n$ are clearly monotone 
decreasing in $r$
for each fixed $n$, and the identity (0.151.1 in Ref.~\cite{GR})
\begin{equation}\label{eq:binom_ident}
\sum_{r=0}^R \binom{q+r}{q} = \binom{q+1+R}{q+1};
\end{equation}
the second line is an elementary algebraic manipulation. A further algebraic manipulation shows that
\begin{equation}
L_{n+1}^{(r)}  =
\frac{p! }{2^{p+1}} 
\frac{r+(n+1)(p+1)}{r+p+1} \binom{(n+1)(p+1)-1}{p}L_n^{(p+1+r)}
\end{equation}
which allows us to conclude that the bound on $K_n^{(r)}$ holds for all $n$ by induction.
Noting that
\begin{equation}
L_n^{(0)} = \frac{(n(p+1))! n^{n}}{ (2^{(p+1)}(p+1))^n (n!)^2}
\end{equation}
we obtain a lower bound on $J_n=K_1^{(0)}K_{n-1}^{(0)}=
p!2^{-(p+1)} K^{(0)}_{n-1}$ as
\begin{equation}\label{eq:Jlower}
J_n  \ge \frac{p!}{2^{p+1}} 
L^{(0)}_{n-1}  = 
\frac{(p+1)! ((n-1)(p+1))! (n-1)^{n-1}}{(2^{p+1} (p+1))^{n}((n-1)!)^2}.
\end{equation}

In a similar way, we find an upper bound
\begin{equation}\label{eq:upper}
K_n^{(r)} \le \binom{n(p+2)-2+r}{n(p+1)-1} U^{(r)}_n
\end{equation}
where
\begin{equation}
U_n^{(r)} = \frac{(n(p+1))!}{n! (2^{p+1}(p+1))^{n}}
\prod_{k=0}^{n-2}\prod_{q=1}^p \frac{k(p+1)+ r + q}{kp + r + n +q -1}
\end{equation}
and the product on $k$ is again regarded as a factor of unity in the case $n=1$.
{}From this expression, it is clear that 
the $U_n^{(r)}$ are monotone increasing in $r$ for each fixed $n$. 
The double product can be also be written as a ratio of products of $\Gamma$-functions and other 
simple functions; in the case $r=0$ there is a particularly simple expression
\begin{equation}
\prod_{k=0}^{n-2}\prod_{q=1}^p \frac{k(p+1) + q}{kp  + n +q -1} = (p+1)^{1-n}.
\end{equation} 
As before, we prove \eqref{eq:upper} by induction, noting that it holds with equality in the case
$n=1$. Supposing that it is true for some $n\ge 1$, the recurrence relation 
Eq.~\eqref{eq:recurrence} gives
\begin{equation}
K_{n+1}^{(r)} \le  \frac{p!}{2^{p+1}} \binom{p+r}{p}\sum_{r'=0}^{p+r+1}
\binom{n(p+2)-2+r'}{n(p+1)-1} U^{(r')}_n.
\end{equation}
Over the summation range, we have $U^{(r')}_n\le U^{(p+1+r)}_n$, so
\begin{equation}
K_{n+1}^{(r)} \le  U^{(p+1+r)}_n\frac{p!}{2^{p+1}} \binom{p+r}{p}\sum_{r'=0}^{p+r+1}
\binom{n(p+2)-2+r'}{n(p+1)-1} \le 
U^{(p+1+r)}_n   \binom{p+r}{p}  \frac{p!}{2^{p+1}} \sum_{r''=0}^{p+r+n}
\binom{n(p+1)-1+r''}{n(p+1)-1},
\end{equation}
where we have changed summation variable to $r''=r'+n-1$ and extended the summation range in the 
second step. Using Eq.~\eqref{eq:binom_ident}, this gives
\begin{equation}
K_{n+1}^{(r)} \le  \frac{p!}{2^{p+1}} \binom{p+r}{p} \binom{n(p+2)+p+r}{n(p+1)}
U^{(p+1+r)}_n 
\end{equation}
Using the fact that
\begin{equation}
\binom{p+r}{p} \binom{n(p+2)+p+r}{n(p+1)} = 
\frac{((n+1)(p+1))!}{(n+1)(p+1)[n(p+1)]! p!}
\binom{(n+1)(p+2)-2+r}{(n+1)(p+1)-1}
\prod_{q=1}^p \frac{r+q}{r+n+q},
\end{equation}
it is then easy to show that \eqref{eq:upper} holds with $n$ replaced by $n+1$ and hence for all $n$ 
by induction. 

We may then obtain the upper bound on $J_n$ as
\begin{equation}
J_n \le \frac{(p+1)!(p+1)^3}{(2^{p+1}(p+1)^2)^n}  
\frac{((n-1)(p+2)-2)!}{((n-2)!)^2} 
\end{equation}
after some manipulation. 

Using Stirling's formula, $(nA-B)!\sim \sqrt{2\pi} (nA/e)^{nA-B+1/2} e^{-B+1/2}$. 
Then one may check that the lower bound in~\eqref{eq:Jlower} is, asymptotically,
\begin{equation}
J_n \gtrsim  \frac{(p+1)!}{2\pi e}\left(\frac{p}{p+1}\right)^{p+1/2} 
\left(\frac{(p+1)^p e}{p^p 2^{p+1}}\right)^n (np-(p+1))!
\end{equation}
while a similar calculation at the upper bound gives
\begin{equation}
J_n \lesssim  \frac{(p+1)! (p+1)^{3}}{2\pi(p+2)^{3}}\left(\frac{p}{p+2}\right)^{p+1/2} 
\left(\frac{(p+2)^{p+2}}{2^{p+1}(p+1)^2 p^p}\right)^n (np-(p+1))!
\end{equation}
so the ratio of the upper bound to the lower bound grows as $\sim \alpha \beta^n$
as $n\to\infty$, 
with 
\[
\alpha  = e \left(\frac{p+1}{p+2}\right)^{p+7/2}
, \qquad
\beta = \frac{1}{e}\left( \frac{p+2}{p+1}\right)^{p+2},
\]
which in the case $p=3$ gives $\alpha = 0.6373520649$, $\beta = 1.122678959$. 
So we have a reasonable control over the leading order contribution. 

As mentioned above, it is certain that the dimensionless moment $a_n$ obeys
 $a_n\ge 8^n J_n$,
and numerical evidence suggests that $a_n\sim 8^n J_n$ at least in the case $p=3$
 (we believe that 
this is true for all $p>1$ and could be proved with more effort). On that basis, we  have
\begin{equation}
 0.513395 \times 3.221667^n  \lesssim \frac{a_n}{(3n-4)!} \lesssim 0.327213\times 3.616898^n
\end{equation}
in the $p=3$ case, for $n\to\infty$. This supports the growth estimates given in the text.

\end{document}